\documentclass[prd,onecolumn,notitlepage,showpacs,preprintnumbers,amsmath,amssymb,nofootinbib,aps,10pt,longbibliography]{revtex4-2}

\pdfoutput = 1

\usepackage{dcolumn}
\usepackage{bm}
\usepackage{xcolor}
\usepackage[utf8]{inputenc}
\usepackage[spanish,english]{babel}
\usepackage{amsmath,amssymb,amsfonts,latexsym,cancel}
\usepackage[normalem]{ulem}
\usepackage{graphicx}
\usepackage{color}
\usepackage{soul}
\usepackage{ulem}
\usepackage[colorlinks=true,linkcolor=red,urlcolor=blue,citecolor=blue]{hyperref}
\usepackage{amsmath}
\usepackage{slashed}
\usepackage{braket}
\usepackage{amssymb}
\usepackage{amsmath}

\usepackage{graphicx}
\allowdisplaybreaks

\newcommand{\be}{\begin{equation}}
\newcommand{\ee}{\end{equation}}
\newcommand{\bea}{\begin{eqnarray}}
\newcommand{\eea}{\end{eqnarray}}
\usepackage{orcidlink,booktabs}
\usepackage{soul,amsmath,amssymb}
\usepackage{siunitx,bm}
\usepackage{comment}

\newcommand{\mi}{\mathrm{i}}
\DeclareMathOperator{\sinc}{sinc}

\renewcommand{\vec}[1]{\boldsymbol{\mathbf{#1}}}
\renewcommand{\unit}[1]{\vec{\hat{#1}}}

\begin{document}
\hspace{5.2in} \mbox{CALT-TH/2024-027}
\title{Signatures of linearized gravity in atom interferometers: %
A simplified computational framework}

\author{Leonardo Badurina}
\email{badurina@caltech.edu}
\author{Yufeng Du}
\email{yfdu@caltech.edu}
\author{Vincent S. H. Lee}
\email{szehiml@caltech.edu}
\author{Yikun Wang}
\email{yikunw@caltech.edu}
\author{Kathryn M. Zurek}
\email{kzurek@caltech.edu}
\affiliation{Walter Burke Institute for Theoretical Physics, California Institute of Technology, Pasadena, CA 91125, USA}
\date{\today}

\begin{abstract}
We develop a general framework for calculating the leading-order, general relativistic contributions to the gravitational phase shift in single-photon atom interferometers within the context of linearized gravity. We show that the atom gradiometer observable, which only depends on the atom interferometer propagation phase, 
can be written
in terms of three distinct contributions: the Doppler phase shift, which accounts for the tidal displacement of atoms along the baseline, the Shapiro phase shift, which accounts for the delay in the arrival time of photons at atom-light interaction points, and the Einstein phase shift, which accounts for the gravitational redshift measured by the atoms. 
For specific atom gradiometer configurations, we derive the signal and response functions for two physically motivated scenarios: \textit{(i)} transient gravitational waves in the transverse-traceless gauge and, for the first time, in the proper detector frame, and \textit{(ii)} transient massive objects sourcing weak and slow-varying Newtonian potentials. We find that the Doppler contribution of realistic Newtonian noise sources (e.g., a freight truck or a piece of space debris) at proposed atom gradiometer experiments, such as AION, MAGIS and AEDGE, can exceed the shot noise level and thus affect physics searches if not properly subtracted.
\end{abstract}

\maketitle

\section{Introduction}
\label{introduction}

Atom interferometry is a versatile and rapidly-developing experimental technique that can be used for a wide variety of precision measurements~\cite{Bongs2019TakingAI}. For instance, atom interferometers (AIs) have been used to measure fundamental constants~\cite{Morelalpha, Estey_2015, doi:10.1126/science.aap7706}, probe the foundational principles of general relativity~\cite{Zych:2011hu, Rosi:2017ieh, Roura:2018cfg, doi:10.1126/science.aay6428, Asenbaum:2020era} and quantum mechanics~\cite{Bassi:2012bg,ManningAI,ArndtAI}, and test models of dark energy and modified gravity~\cite{Burrage:2014oza, Hamilton:2015zga, Elder:2016yxm, Sabulsky:2018jma}. 
Atom gradiometers (AGs), which consist of two spatially separated AIs that are referenced by common lasers, have also been proposed to detect gravitational waves (GWs) in the unexplored ``mid-frequency band''~\cite{Dimopoulos:2007cj, Graham:2012sy, Ellis:2020lxl, Badurina:2021rgt,Banks:2023eym,Baum:2023rwc}, search for violations of the universality of free fall~\cite{Graham:2015ifn} and measure time-varying corrections to atomic transition energies induced by scalar ultralight dark matter~\cite{Arvanitaki:2016fyj, Badurina:2021lwr, Badurina:2023wpk}.

In recent years, a number of ambitious AG experiments have been proposed as quantum sensors for fundamental physics (see Ref.~\cite{Proceedings:2023mkp} for a recent review). These include large-scale terrestrial experiments, such as AION~\cite{Badurina:2019hst}, MAGIS~\cite{MAGIS-100:2021etm}, MIGA~\cite{Canuel:2017rrp}, ELGAR~\cite{Canuel:2019abg}, and ZAIGA~\cite{Zhan:2019quq}, and futuristic space-based experiments, such as STE-QUEST~\cite{STE-QUEST:2022eww} and AEDGE~\cite{AEDGE:2019nxb}. By overcoming a number of experimental systematics, these experiments are expected to operate at the shot-noise level. However, fluctuations in an atom's local gravitational field, e.g., due to seismic waves \cite{Badurina:2022ngn} or Newtonian noise (NN)~\cite{Carlton:2023ffl}, may significantly reduce the projected reach in the key $(10^{-3}-1)$~Hz frequency window. If left unmitigated, these effects will dramatically limit the physics potential of these ambitious experiments. 

Instead of cutting large sections of an experiment's time series~\cite{Carlton:2023ffl},
it may be possible to subtract the phase shift from transient sources of NN and recover an experiment's shot-noise limited sensitivity \textit{provided} that this phase shift is known with sufficient precision. Depending on an experiment's projected reach and type of physics search, such a strategy may require a fully relativistic calculation. For example, a massive object traveling with speed $v_s$ will induce a phase shift by accelerating the atoms, given by $\Delta \phi_\mathrm{non-rel} \sim k_\mathrm{eff} a T^2$~\cite{PhysRevLett.67.181}, with $k_\mathrm{eff}$ being the maximum momentum difference between the two arms of an interferometer, $T$ being the interrogation time, and $a$ being the acceleration of the atoms. This is typically calculated utilizing only classical Newtonian mechanics. However, relativistic corrections are expected. %
Using dimensional analysis and the invariance of general relativity under parity and time-reversal, the dominant relativistic phase shift is expected to scale as $\Delta \phi_\mathrm{rel} \sim v_s \Delta \phi_\mathrm{non-rel}$. Although useful to estimate the size of the effect, $\Delta \phi_\mathrm{rel}$ does \textit{not} inform us of: \textit{(i)} the coefficient of this phase shift term, which may be sequence dependent, and \textit{(ii)} the shape of the power spectrum associated with this phase shift, both of which can only be inferred from a general-relativistic (i.e., general coordinate-invariant) phase shift calculation. 

Several coordinate-invariant formalisms have been proposed~\cite{Dimopoulos:2006nk,Dimopoulos:2008hx,Roura:2018cfg,Werner:2023qfu,Roura:2024rmm} and have been used to calculate, e.g., the phase shift induced by gravitational waves~\cite{Dimopoulos:2007cj,Graham:2012sy,Graham:2016plp} and static potentials (e.g., the Earth's gravitational field)~\cite{Dimopoulos:2006nk,Roura:2018cfg,Werner:2023qfu,Roura:2024rmm}. Notably, the formalism proposed in Ref.~\cite{Dimopoulos:2006nk} defines all tunable experimental parameters in a frame independent manner, solves for the geodesics of the freely falling atomic wavepackets and laser pulses, and determines the momentum transferred to the atoms as a result of atom-light interactions in the atom's local interial frame. Although crucial for correctly predicting the size of relativistic contributions to the phase shift, these formalisms are computationally cumbersome when the number of interaction points exceeds $\mathcal{O}(1)$.~This is especially relevant in proposed atom gradiometer experiments employing large momentum transfer (LMT) such as AION and MAGIS, which plan up to $\mathcal{O}(10^4)$ atom-light interactions per cycle.%

Aside from these computational considerations, outstanding questions remain about the physical interpretation of existing gauge-invariant frameworks for computing gravitational phase shifts, such as the interpretation of AGs as gravitational antennas, or equivalently the mapping between AG and laser interferometer observables.
For example, Ref.~\cite{Dimopoulos:2006nk} decomposes the gauge-invariant phase shift for a single AI (and consequently for an AG) into three contributions: the phase shift associated with the free evolution of atomic wavepackets in spacetime (i.e., the propagation phase), the phase shift imprinted by the laser pulses during atom-light interactions (i.e., the laser phase) and the phase shift associated with the degree to which the two spatially separated wavepackets do not overlap at the application of the final beamsplitter pulse (i.e., the separation phase). As shown for GWs in Ref.~\cite{Rakhmanov:2004eh} and recently for more general metric perturbations via a proper time treatment in Ref.~\cite{Lee:2024oxo}, the observable in laser interferometers can be written as a sum of three distinct (and separately not diffeomorphism-invariant) contributions: the time delay caused by the tidal displacement of the mirrors along the baseline (i.e. the Doppler time delay), the delay in the arrival time of photons at the mirrors (i.e. the Shapiro time delay), and the time delay due to the gravitational redshift measured by the beamsplitter  (i.e. the Einstein time delay). Since AGs have been proposed as gravitational wave interferometers, it should be possible to extract these three contributions from the gauge-invariant AG phase shift. 
This endeavor would clarify the interpretation of AGs as exquisite accelerometers~\cite{RevModPhys811051} and time-keeping devices~\cite{RevModPhys87637}, and elucidate the origin of the relativistic phase shift contributions in different frames.

In this work, we address these points by developing from first principles a simplified 
general coordinate-invariant framework for calculating phase shifts in single-photon AGs.~Since the metric perturbation in the problems of interest is very small, we will work within the context of linearized gravity.
Additionally, as the motion of the atoms relative to the laser sources is highly non-relativistic, we work to leading order in the atom velocity.
Notably, for the case of AGs, we express the differential phase shift in terms of contributions which are in one-to-one correspondence with the time delays that enter the laser interferometer observable: the phase shifts associated with the Doppler, Shapiro and Einstein time delays.
Equipped with this formalism, we compute the signal and response function induced by two well-motivated physical scenarios: transient GWs and weak and slow-varying Newtonian potentials sourced by transient massive objects. Since the form of the response function depends on the pulse sequence, we will perform explicit calculations for gradiometers employing Mach-Zehnder and LMT configurations. Importantly, our examples highlight the gauge-invariance of our framework (as explicitly shown in the GW calculation, which is performed in both the transverse-traceless and the proper detector frame) and the accuracy of our formalism in reproducing existing results in the literature in a more physically and computationally transparent fashion.

This paper is structured as follows. After reviewing the basics of atom interferometry, in section~\ref{sec:framework} we introduce our general coordinate-invariant framework for computing gradiometer phase shifts in linearized gravity. After deriving the basis for our formalism in section~\ref{sec:framework_derivation}, in section~\ref{sec:framework_derivation:definite_pulses} we introduce the gradiometer observable and provide expressions for Mach-Zehnder and LMT gradiometer configurations.
As example applications, in section~\ref{sec:applications} we compute the phase shifts induced by gravitational waves and slow-varying weak Newtonian potentials. In section~\ref{sec:disc_conc} we summarize the key results of this paper. Appendices~\ref{app:sep_phase}--\ref{sec:geodesic} support the calculations in sections~\ref{sec:framework}--\ref{sec:applications}.

\section{Derivation of the Leading-order Gradiometer Phase Shift in Linearized Gravity}\label{sec:framework}

Schematically, atom interferometers (AIs) utilize matter wave interference to detect the phase difference between two coherent atomic states in a spatial superposition. In order for an atom to be prepared in a spatial superposition and then measured via matter wave interference, the trajectories of the atomic wavepackets are manipulated using laser pulses. Most experiments rely on a two-level system where the external momentum and the internal energy state of the atoms can be manipulated via Rabi oscillations \cite{Abend:2020djo}. A laser pulse that interacts with the atom over a quarter of the Rabi cycle (a $\pi/2$-pulse) takes an atom in one state to an equal superposition of two states, thus acting as a ``beamsplitter"; a pulse over half of the Rabi cycle (a $\pi$-pulse) reverses the state of the atom, thus acting as a ``mirror"~\cite{foot2005atomic}. In this paper, we focus on experiments that rely on two spatially separated AIs operating common lasers and single-photon transitions with energy separation $\omega_a$. We refer to these configurations as single-photon atom gradiometers (AGs). 

In order to correctly capture relativistic effects, which manifest through the dependence of dynamics on the curvature of spacetime, it is of paramount importance to describe the interferometer sequence and all experimental quantities in a frame-independent manner. 
Importantly, this guarantees that the observable is free of gauge artifacts. Let us consider the description of atomic fountains (which we assume in this work) in the language of general relativity (e.g., Ref.~\cite{Dimopoulos:2006nk}). In these AI experiments, the atoms are in free fall. Provided that the radius of curvature is much larger than the wavepacket size, a semi-classical treatment is sufficient. In this regime, atom trajectories are described in terms of timelike geodesics, while photon trajectories are described in terms of null geodesics. The spacetime points at which atom-light interactions occur can be solved in coordinate time and position. Furthermore, for the phase shift to be frame-independent,  %
the coordinate times at which the laser pulses are emitted must be related to the established time difference measured %
by an observer traveling along the laser's worldline. Finally, the four momentum of a pulse must be related to the frequency of the pulse at emission; this four momentum is then evolved from emission to the designated atom-light interaction point, where the atom's recoil is computed in a local inertial frame.

Inspired by this description, in the following sections we provide a detailed derivation from first principles of the gauge-invariant \textit{gradiometer} phase shift. Importantly, our framework is valid to leading order in a generic metric perturbation and correctly predicts the coefficients of phase shift terms that are linear in the atom's recoil velocity. Starting from the phase shift for a single AI, we show that the single-photon AG observable can be reinterpreted in terms of coordinate time delays that are well understood in the context of laser interferometers: the Doppler, Shapiro and Einstein time delays. 

\subsection{Propagation phase shift}
\label{sec:framework_derivation}

Let us consider the semi-classical evolution of an atomic wavepacket $\ket{\psi}$ in spacetime. In this regime, the atom's dynamics can be described in terms of the evolution of the atom's center-of-mass (c.o.m.). Consequently, solving the Schr{\"o}dinger equation, $\ket{\psi} \propto \exp(\mi S)$, where $S$ is the action of the atom's c.o.m.~\cite{Sakurai:2011zz}. In general relativity,
the action of a massive point-like particle can be expressed in terms of the proper time elapsed along particle's timelike worldline~\cite{Misner:1973prb}.
Therefore, the phase difference associated with the spacetime propagation of %
two spatially separated wavepackets corresponds to the \textit{difference} in the actions evaluated along the worldlines of the two wavepackets' c.o.m. from state initialization to measurement. For a spacetime with metric $g_{\mu \nu}$, the propagation phase is
\begin{equation}\label{eq:prop_unperturbed}
    \Delta \phi =  \oint_{C} m d\tau = \oint_{C} m \sqrt{-g_{\mu \nu} \frac{dx^\mu}{dt} \frac{dx^\nu}{dt}} dt \, ,%
\end{equation}
where $m$ is the path-dependent mass of the atom, with $m = m_o$ and $m = m_o + \omega_a$ for ground and excited states, respectively.\footnote{We work in natural units (i.e. $\hbar = c = 1$) and we use the convention $\eta_{\mu\nu} = \mathrm{diag}(-1,+1,+1,+1)$ for the flat spacetime metric.} The proper time along the atom's worldline is defined as $\tau$, which is not to be confused with the coordinate time $x^0 = t$, and the  atom's path-dependent four coordinate velocity is defined as $dx^\mu/dt$. We parameterize the atom's evolution with respect to $t$ and perform the loop integral over the closed semi-classical path $C$. This path depends on the following: \textit{(i)} the free evolution of the atomic wavepackets between atom-light interaction points, \textit{(ii)} the arrival time of laser pulses at atom-light interaction points and \textit{(iii)} the metric-dependent correction to the laser beam's wave-vector, which leads to a correction in the recoil of the atoms after atom-light interactions. Here, we assume that $C$ is initiated when the superposition of states is created (i.e. at the initial beamsplitter pulse) and closes when the interference pattern is measured. The latter should not be confused with the final beamsplitter used to redirect the atomic states to the measurement ports. 

Note that the full AI phase observable, $\Delta \phi \big |_{\mathrm{AI}}$, also depends on the laser phase $\Delta \phi_\mathrm{laser}$, which is imprinted onto the atoms during atom-light interaction and arises from the linear coupling between the photon field and the atom's electric dipole~\cite{foot2005atomic}:
\begin{equation}\label{eq:AIobservable}
    \Delta \phi \big |_{\mathrm{AI}}  =  \Delta \phi + \Delta \phi_\mathrm{laser}. \,
\end{equation}
In gradiometer setups featuring more than one AI referenced by the same laser, the AIs experience a common laser phase which cancels out in a differential measurement.\footnote{Since the radius of curvature considered in this work is much larger than the wavelength of the electromagnetic waves driving the Rabi oscillations, it suffices to work in the limit of geometric optics. In this limit, the phase of electromagnetic waves is constant along null geodesics and can be defined with respect to any timelike observer~\cite{Misner:1973prb,Wald:1984rg}. For instance, Ref.~\cite{Dimopoulos:2006nk} defines this phase with respect to an observer traveling along the laser's worldline.} Since we focus on gradiometer observables, we do not consider this effect further, though it is important to note that the laser phase contribution is generally non-zero in single-AI setups.

\begin{figure}[t]
	\includegraphics[width=0.35 \textwidth]{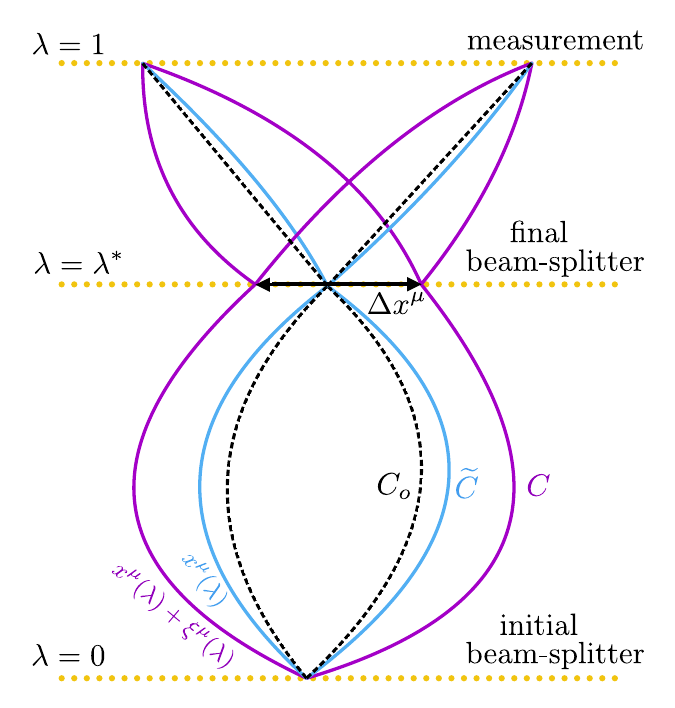} 
	\caption{Schematic spacetime diagram showing the atom trajectories used in deriving the framework. The dotted yellow lines denote hypersurfaces defined by the initial beam-splitter pulse, the final beam-splitter pulse, and the measurement. The unperturbed atom trajectory $C_o$ is schematically shown with dashed black lines. In the presence of a metric perturbation, the atom follows a geodesic $C$ (purple), which can be deformed into another geodesic $\widetilde{C}$ (blue) that closes at the final beam-splitter. $\Delta x^\mu$ is the separation of path $C$ at the final beam-splitter pulse. $\lambda$ parameterizes the atom geodesics, and the deformation $\xi(\lambda)$ is of order $\mathcal{O}(h)$ and only enters into the observable at $\mathcal{O}(h^2)$. At $\mathcal{O}(hv)$, $\widetilde{C}$ is a good approximation for $C$ in calculating the propagation phase shift.}\label{fig:path_def}
\end{figure}

For the relativistic effects considered in this work, it is sufficient to linearize the metric tensor, i.e. $g_{\mu\nu} = \eta_{\mu \nu} + h_{\mu \nu}$, with $h_{\mu\nu}\ll 1$. By working at leading order in $h_{\mu \nu}$ and relating $C$ to ancillary paths, which we schematically depict in Fig.~\ref{fig:path_def}, a number of simplifications will arise. First, in the absence of metric perturbations, $g_{\mu \nu} = \eta_{\mu \nu}$, we assume that the atomic wavepackets travel along trajectories that recombine at the application of the final beamsplitter pulse; we will refer to this path as $C_o$. Second, as alluded to above, we define $C$ as the closed path traced by the atomic wavepackets in the presence of a non-zero metric perturbation $h_{\mu \nu}$; 
however, while $C$ is closed at measurement, it is {\it no longer closed} at the final beamsplitter pulse. Finally, we define a third and fictitious path $\widetilde{C}$, constructed from $C$ assuming the same metric perturbation $h_{\mu \nu}$; the initial conditions (e.g.~atom velocity) are adjusted for the path to {\it close} at the final beamsplitter pulse. Hence, generally $\widetilde{C} \neq C$.

The advantage of introducing these ancillary paths is that, at leading order in the metric perturbation, the phase shift along the true path $C$ is equivalent to the phase shift computed along the unphysical path $\widetilde{C}$.  This can be seen simply as follows. Let $\xi^\mu(\lambda)$ be the coordinate separation between the atom geodesics in $C$ and $\widetilde{C}$, where $\lambda \in [0,1]$ parameterizes the atom worldlines and $\xi^\mu(0) = \xi^\mu(1) = 0$, since both $C$ and $\widetilde{C}$ are closed with respect to state initialization and state measurement. 
Expanding in the coordinate deviation $\xi$ and defining $L = m\sqrt{g_{\mu\nu} \dot x^\mu \dot x^\nu}$, where $\dot x^\mu = d x^\mu/d\lambda$, the phase shift over the true path $C$ takes the form\footnote{
Note that, while we assume here that $C$ is an exactly closed path {\it at measurement} in the presence of metric perturbations, $C$ is generally not precisely closed even at measurement.  This leads to a correction to the phase shift which appears at leading order in the metric perturbation $h$ and $v$. Nevertheless, as discussed in Appendix~\ref{app:sep_phase}, the separation phase can be neglected by choosing an appropriate ancillary path.}
\begin{widetext}
\begin{equation}
\begin{aligned}
\Delta \phi & =
\oint_{C}m d\tau \\ & = \oint_{\widetilde{C}} m d\tau 
+ 
\int_{0}^1  \xi^\mu \left( \frac{d}{dt }\frac{\partial L}{\partial \dot x^\mu} - \frac{\partial L}{\partial x^\mu} \right)\Bigg |_{\mathcal{R}} d\lambda - \int_{0}^1 \xi^\mu \left( \frac{d}{dt }\frac{\partial L}{\partial \dot x^\mu} - \frac{\partial L}{\partial x^\mu} \right)\Bigg |_{\mathcal{L}} d\lambda 
+\mathcal{O}(\xi^2) \, \\
 & = \oint_{\widetilde{C}} m d\tau +\mathcal{O}(\xi^2) \, ,
\end{aligned}
\label{eq:CtildetoC}
\end{equation}
where the left and right trajectories are marked by the $\mathcal{L}$ and $\mathcal{R}$ subscripts, respectively.
The second and third terms in the second line vanish because the atom worldlines in $\widetilde{C}$ are geodesics, and hence extrema of the action.  

We next expand Eq.~\eqref{eq:CtildetoC} to leading order in the metric perturbation, utilizing $\widetilde{C} = C_o + \delta C$ with $\delta C = \mathcal{O}(h)$:
\begin{equation}\label{eq:phi_tot_1}
\begin{aligned}
    \Delta \phi &= \oint_{\widetilde{C}} m \sqrt{-\eta_{\mu\nu}\frac{dx^\mu}{dt} \frac{dx^\nu}{dt}} dt 
    -  \frac{1}{2} \oint_{C_o} \frac{m}{\sqrt{-\eta_{\mu\nu}\frac{dx^\mu}{dt} \frac{dx^\nu}{dt}}} \,  h_{\mu\nu}\frac{dx^\mu}{dt} \frac{dx^\nu}{dt} dt 
    + \mathcal{O}(h^2) \, . 
    \end{aligned}
\end{equation}
Writing the phase in Eq.~\eqref{eq:phi_tot_1} to leading order in $\mathcal{O}(hv)$, we find%
\footnote{Although this approximation is frame dependent, we can always choose to work in a frame where $v\ll1$. The only possible transformation from a frame where the atoms are non-relativistic to one where the atoms are moving at relativistic speeds, and which does not spoil the gauge transformation in the metric in linearized gravity, is a Lorentz transformation. Such a transformation is global and therefore boosts the entire laboratory apparatus, namely the atoms and the laser sources.}
\begin{equation}\label{eq:phi_tot_first_order1}
\begin{aligned}
    \Delta \phi &= \oint_{\widetilde{C}} m dt 
    -\oint_{C_o} m v^i\delta v_i dt - \frac{1}{2} \oint_{C_o} m h_{00} dt - \oint_{C_o} m h_{0i}v^i dt 
    +\mathcal{O}(h^2,v^2) \, ,
\end{aligned}
\end{equation}
where $\delta v_i$ is the $\mathcal{O}(h)$ correction to the atom three velocity arising from free-falling atoms, obtained by solving geodesic equations\footnote{Recall that the coordinate four velocity is defined as $dx^\mu/dt$, where $t$ is coordinate time. Hence, metric perturbations can only affect the $\mu \in \{1,2,3\}$ components of this four vector, $dx^\mu/dt = (1, v^i + \delta v^i)$, where $\delta v^i = \mathcal{O}(h)$.} 
\begin{equation}\label{eq:deltav}
\begin{aligned}
   \delta v^i(t,x) &=  -\int_{t_0}^{t} dt' \, \Gamma^i_{00}(x,t') 
   = - \eta^{ij}\int_{t_0}^t dt' \, \left (\partial_0 h_{j0} -\frac{1}{2} \partial_j h_{00} \right ) 
   + \mathcal{O}(h^2,v)\, ,
\end{aligned}
\end{equation}
\end{widetext}
where $\Gamma^\mu_{\alpha \beta}$ is the Christoffel symbol in linearized gravity and $t_0$ is the time at which the experiment is initialized. We neglect the contribution to the atom's recoil velocity after atom-light interactions that arises from the $\mathcal{O}(h)$ correction to the photon's wave-vector, since the leading-order correction is proportional to $v$ and therefore enters the observable at next-to-leading order.

The first term on the RHS of Eq.~\eqref{eq:phi_tot_first_order1} corresponds to phase shift due to the deformation of the atom trajectories in going from $C_o$ to $\widetilde{C}$. Because we are interested in signatures of linearized gravity, we only consider $\mathcal{O}(h)$ effects, which are attributed to $\delta C$. Since multiple pulses divide each interferometer arm into path segments, which we label by $k$, the first term on the RHS of Eq.~\eqref{eq:phi_tot_first_order1} can be written as
\begin{equation} \label{eq:SD_phase} 
\begin{aligned}
\oint_{\widetilde{C}}m dt & \supset \oint_{\delta C}m dt 
= \left(\sum_{k\in E} \omega_a \Delta t^{(k)}\right)\Bigg|_{\mathcal{R}} - \left(\sum_{k\in E} \omega_a \Delta t^{(k)}\right)\Bigg|_{\mathcal{L}} \, .
\end{aligned}
\end{equation}
Notice that the leading order term is proportional to $\omega_a$ since the rest mass contribution vanishes under a loop-integral, i.e. $\oint_{\widetilde{C}}m_odt=0$;\footnote{In symmetric configurations, $\oint_{C_o}mdt=0$, so that $\oint_{\widetilde{C}} mdt=\oint_{\delta C}mdt$.} hence we only sum over the excited segments, denoted as the set $E$.  As a result, only excited state path segments contribute to the phase shift in~Eq.~\eqref{eq:SD_phase}; this will be very important in the calculations below. In Eq.~\eqref{eq:SD_phase}, $\Delta t^{(k)}$ is the $\mathcal{O}(h)$ perturbation to the coordinate time duration of the $k$-th path segment, and path segments on both the right ($\mathcal{R}$) and left arm ($\mathcal{L}$) are summed over. The overall minus sign in the left arm contributions originates from the loop integral. 
The coordinate time corrections can be computed by solving for the intersection of atom and photon worldlines. Importantly, these atom-light interaction points define the path segments. We label each path segment $k$ by the laser pulse that starts the sequence. The $k$-th segment is ended by the laser pulse labeled as $k+1$, which starts the subsequent segment $k+1$. Denoting the perturbed initial and final times of the $k$-th path segment as $\bar{t}^{(k)}+\delta t^{(k)}$ and $\bar{t}^{(k+1)}+\delta t^{(k+1)}$, the perturbed atom worldline as $\bar{x}^i(t)+\delta x^i(t)$, the worldlines of the photons as $\bar{x}^i_{\gamma^{(k)}}(t)+\delta x^i_{\gamma^{(k)}}(t)$ and $\bar{x}^i_{\gamma^{(k+1)}}(t)+\delta x^i_{\gamma^{(k+1)}}(t)$, 
with unperturbed quantities denoted by overbars, and $n^i$ as the unit vector pointing along the baseline (considering the motion of both the photons and atoms in a single spatial direction), the corresponding equation to solve is
\begin{equation}\label{eq:worldline_intersection}
\begin{aligned}
   n_i & \left[\bar{x}^i \left (\bar{t}^{(k)}+\delta t^{(k)}\right )+\delta x^i \left (\bar{t}^{(k)} \right )\right] = 
   n_i\left[\bar{x}_{\gamma^{(k)}}^i \left (\bar{t}^{(k)}+\delta t^{(k)}\right )+\delta x_{\gamma^{(k)}}^i \left (\bar{t}^{(k)} \right )\right] \, .
\end{aligned}
\end{equation}
Expanding Eq.~\eqref{eq:worldline_intersection} to $\mathcal{O}(h)$ and neglecting the unperturbed atom velocity (as the prefactor $\omega_a$ in Eq.~\eqref{eq:SD_phase} is parametrically $\mathcal{O}(v)$), we find
\begin{equation}\label{eq:worldline_intersection_2}
    \delta t^{(k)} = (\pm)_{(k)} n_i\left[\delta x^i \left (t^{(k)} \right ) - \delta x^i_{\gamma^{(k)}}\left (t^{(k)} \right)\right] \, ,
\end{equation}
where $(\pm)_{(k)}$ is taken to be $+1$ or $-1$ for outgoing or incoming photons (i.e. parallel or anti-parallel to the baseline) interacting with the atom, respectively. Here we drop the overbars, which is valid to $\mathcal{O}(h)$. The perturbed atom positions, $\delta x^i(t)$, are given by integrating Eq.~\eqref{eq:deltav}, i.e.
\begin{equation}\label{eq:delta_x}
    \delta x^i(t,x) = \int_{t_0}^{t}\delta v^i(t',x) \, dt' \, ,
\end{equation}
and the perturbed photon trajectories, $\delta x_{\gamma^{(k)}}^i(t)$, are given by solving the null geodesic condition, $ds^2=g_{\mu\nu}dx^{\mu}dx^{\nu}=0$, leading to
\begin{equation}\label{eq:delta_x_gamma}
    n_i \frac{d}{dt}\delta x^i_{\gamma^{(k)}}(t) = (\mp)_{(k)} \mathcal{H}_{(\pm)_{(k)}}\, ,
\end{equation}
where we defined
\begin{equation}\label{eq:mathcal_H}
    \mathcal{H}_{\pm}\equiv \frac{1}{2}\left(h_{00}\pm 2h_{0i}n^i+h_{ij}n^in^j\right) \, .
\end{equation}
Putting Eq.~\eqref{eq:delta_x} and Eq.~\eqref{eq:delta_x_gamma} into Eq.~\eqref{eq:worldline_intersection_2}, 
the $\mathcal{O}(h)$ perturbation to the coordinate time duration of the $k$-th path segment, namely $\Delta t^{(k)}$, can thus be expressed as the sum of Doppler and Shapiro time delays, %
\begin{align}\label{eq:dt_Doppler_Shapiro}
   \Delta t^{(k)} = \delta t^{(k+1)} - \delta t^{(k)} \equiv \Delta t_{\mathcal{D}}^{(k)} +\Delta t_{\mathcal{S}}^{(k)} \, .
\end{align}
The Doppler term is due to the atom's motion under the metric perturbation as derived in Eq.~\eqref{eq:delta_x}:
\begin{equation}\label{eq:Doppler_schematic}
\begin{aligned}
    \Delta t^{(k)}_{\mathcal{D}}
     & = n_i\Big[(\pm)_{(k+1)}\delta x^{i}\left (t^{(k+1)},x^{(k+1)}\right ) 
     - (\pm)_{(k)}\delta x^{i}\left (t^{(k)},x^{(k)}\right)\Big] \, ,
\end{aligned}
\end{equation}
with $x^{(k)}$ and $x^{(k+1)}$ being the unperturbed initial and final atom positions of the $k$-th and $(k+1)$-th path segment. The Shapiro term, which corresponds to the time delay accrued by the photon along its geodesic as derived in Eq.~\eqref{eq:delta_x_gamma}, is given by 
\begin{equation}\label{eq:Shapiro_schematic}
\begin{aligned}
    \Delta t^{(k)}_{\mathcal{S}} & = \Delta\mathcal{T}^{(\pm)_{(k+1)}}_\mathcal{S}\left (t_{\mathrm{L}}^{(k+1)}, x_{\mathrm{L}}^{(k+1)}, x^{(k+1)}\right) 
    -\Delta\mathcal{T}^{(\pm)_{(k)}}_\mathcal{S}\left(t_{\mathrm{L}}^{(k)}, x_{\mathrm{L}}^{(k)}, x^{(k)} \right) + 
        \Delta t_\mathrm{laser}^{(k)} \, ,
\end{aligned}
\end{equation}
where we defined
\begin{equation}\label{eq:TScurl}
    \Delta \mathcal{T}_\mathcal{S}^\pm \left (t, x_1, x_2 \right)\equiv \pm\int_{x_1}^{x_2}  \mathcal{H}_{\pm}\left (t\pm(x'-x_1), x'\right )dx' \, .
\end{equation}
Here, the photons which interact with the atom at the $k$-th and $(k+1)$-th intersection are emitted from the lasers at $x_{\mathrm{L}}^{(k)}$ and $x_{\mathrm{L}}^{(k+1)}$, and $t^{(k)}-t_{\mathrm{L}}^{(k)}\equiv  (\pm)_{(k)}\left (x^{(k)}-x_{\mathrm{L}}^{(k)}\right )$ %
with $t_\mathrm{L}^{(k)}$ the corresponding photon emission time. The quantity $\Delta t_\mathrm{laser}^{(k)}$ depends on the $\mathcal{O}(h)$ correction to the photon's emission spacetime points and cancels out in a differential measurement, as discussed in Appendix~\ref{app:DopplerShapiro}.

The remaining three terms on the RHS of Eq.~\eqref{eq:phi_tot_first_order1} only depend on the unperturbed path $C_o$. By using Eq.~\eqref{eq:deltav}, these terms can be rewritten as
\begin{equation}
\begin{aligned} \label{eq:E0_1}
\Delta \phi_\mathcal{E} & \equiv  -\oint_{C_o} m v^i\delta v_i dt - \frac{1}{2} \oint_{C_o} m h_{00} dt - \oint_{C_o} m h_{0i}v^i dt \\ & = - \frac{1}{2} \oint_{C_o}dt\, m \left (1 + v^i \int^t_{t_0} dt' \partial_i \right ) h_{00}   \, .
\end{aligned}
\end{equation}
These contributions only depend on $h_{00}$, and are thus identified as the Einstein term (i.e. time dilation as measured by the atoms). %
We note that this phase shift can be significantly simplified if the metric perturbation is spatially slow varying
over $C_o$, which allows it to be expanded as
\begin{equation}\label{eq:long_wavelength}
\begin{aligned}
    h_{\mu\nu}(t,{x}_\mathrm{AI}+ {x}_a)
    & =
    h_{\mu\nu}(t,{x}_\mathrm{AI})+x_{a}^i \partial_i h_{\mu\nu}(t,{x}_\mathrm{AI}) 
    + \mathcal{O}(v^2) \, .
\end{aligned}
\end{equation}
Here, the distance traveled by the atoms from their initial position is $x_a^i(t)\propto v^i$ and $x_\mathrm{AI}$ is the unperturbed position of the AI at the start of the sequence. The expansion is valid when $(vT)\partial_i h_{\mu\nu}\ll~h_{\mu\nu}$, where $T$ is the interrogation time, and holds for all examples considered in this work.\footnote{As an example, for gravitational waves with angular frequency $\omega$, the condition is satisfied when $\omega \ll (v_aT)^{-1} \sim 1\,\mathrm{GHz}\,(10^{-9}/v_a)(1\,\mathrm{s}/T)$. This requirement is met by many orders of magnitude, as most proposed AI experiments are not sensitive to gravitational waves beyond $1$ Hz.} With this approximation and integrating by parts, Eq.~\eqref{eq:E0_1} simplifies to 
\begin{equation}\label{eq:phaseE}
\Delta\phi_\mathcal{E}= \left(\sum_{k\in E} \omega_a \Delta t_{\mathcal{E}}^{(k)}\right)\Bigg|_{\mathcal{R}} - \left(\sum_{k\in E} \omega_a \Delta t_{\mathcal{E}}^{(k)}\right)\Bigg|_{\mathcal{L}} \, ,
\end{equation}
where the non-vanishing contribution comes from the excited state segments, as the ground state contribution sums to zero over the closed loop, and
\begin{equation}\label{eq:tE0}
 \Delta t^{(k)}_\mathcal{E} = -\frac{1}{2} \int_{C_o^{(k)}}h_{00}(t,{x}_\mathrm{AI})dt + \mathcal{O}(v^2) \, 
\end{equation}
defines the Einstein time delay.

In summary, the propagation phase shift in Eq.~\eqref{eq:phi_tot_first_order1} can be written schematically as
\begin{equation}\label{eq:dse}
\Delta \phi = \Delta \phi_{\mathcal{D}}  + \Delta \phi_{\mathcal{S}} + \Delta \phi_{\mathcal{E}} +\mathcal{O}(h^2, v^2) \, ,
\end{equation}
where the Doppler and Shapiro phase shifts originate from Eq.~\eqref{eq:SD_phase} with $\Delta t^{(k)}$ given in Eq.~\eqref{eq:dt_Doppler_Shapiro}, while the Einstein phase shift originates from Eq.~\eqref{eq:E0_1}. The expressions of the Doppler, Shapiro and Einstein time delays are given in Eqs.~\eqref{eq:Doppler_schematic},~\eqref{eq:Shapiro_schematic} and~\eqref{eq:tE0}, respectively. Together with Eq.~\eqref{eq:phaseE}, Eq.~\eqref{eq:dse} %
facilitates the direct comparison between atom interferometer experiments, such as AION and MAGIS, and laser interferometers, such as LIGO~\cite{LIGOScientific:2007fwp} and GQuEST~\cite{Vermeulen:2024vgl}. Indeed, the time delays that enter the atom interferometer propagation phase also appear in laser interferometer calculations. The only difference between the two propagation phase shifts is the physical origin of the frequency: in the case of atom interferometers, the phase shift is proportional to the energy difference between the ground and excited state; in the case of laser interferometers, the phase shift is proportional to the frequency of the laser pulse traveling along the baseline. 
More broadly, the appearance of the Doppler and Shapiro time delays facilitates the analogy between the atom interferometers and the mirrors of a one-dimensional laser interferometer. Furthermore, the appearance of the Einstein time delay clarifies the analogy between an atom interferometer and the beamsplitter in a laser interferometer. Finally, since we started with the action for massive point-like particles (\textit{cf}.~Eq.~\eqref{eq:prop_unperturbed}), the phase shift is manifestly invariant under diffeomorphisms.

\subsection{Doppler, Shapiro, and Einstein phase shifts for atom gradiometers}
\label{sec:framework_derivation:definite_pulses}

In an AG, the observable is the difference between the phase shifts measured by each AI, i.e. $\Delta \phi_\text{grad}\equiv \Delta\phi\big |_{\mathrm{AI}_1}-\Delta\phi\big |_{\mathrm{AI}_2}$. Since the transitions in each AI are driven by common laser pulses, the gradiometer observable exclusively depends on the difference between the propagation phase shifts of each AI and, subsequently, on the difference between the coordinate time corrections between path segments, which we refer to as gradiometer time delays. 
Without loss of generality, we can write the gradiometer time delays for each segment $k$ because the pair of laser pulses that start and end the path segment in a given AI also start and end a path segment in the other AI. %
Restricting our attention to such configurations, %
the $\mathcal{O}(h)$ correction to the laser's motion cancels, as we carefully show in Appendix~\ref{app:DopplerShapiro}. %
Pulling all parts of the derivation together, the gradiometer phase shift for experiments using single-photon transitions can be schematically expressed as %
\begin{widetext}
\begin{equation}\label{eq:grad-form}
\begin{gathered}
\Delta \phi_\mathrm{grad} = \Delta \phi_{\mathrm{grad},\mathcal{D}}  + \Delta \phi_{\mathrm{grad},\mathcal{S}} + \Delta \phi_{\mathrm{grad},\mathcal{E}} +\mathcal{O}(h^2, v^2) \, , \\
\Delta \phi_{\mathrm{grad}, \mathcal{D,S,E}} = \sum_{k} \Delta \phi_{\mathrm{grad}, \mathcal{D,S,E}}^{(k)} =  \left(\sum_{k\in E} \omega_a \Delta t_{\mathrm{grad},\mathcal{D,S,E}}^{(k)}\right)\Bigg|_{\mathcal{R}} - \left(\sum_{k\in E} \omega_a \Delta t_{\mathrm{grad},\mathcal{D,S,E}}^{(k)}\right)\Bigg|_{\mathcal{L}} \, ,
\end{gathered}
\end{equation}
\end{widetext}
where the gradiometer time delay is expressed in terms of the single AI time delays $\Delta t_{\mathcal{D,S,E}}^{(k)}$ (cf. Eqs.~\eqref{eq:Doppler_schematic},~\eqref{eq:Shapiro_schematic} and \eqref{eq:tE0}) as
\begin{equation}
    \Delta t^{(k)}_{\mathrm{grad},\mathcal{D,S,E}} \equiv \Delta t^{(k)}_\mathcal{D,S,E}\Big |_{\mathrm{AI}_1}-\Delta t^{(k)}_\mathcal{D,S,E}\Big |_{\mathrm{AI}_2} \, .
\end{equation}
Since the ground state contribution vanishes over the loop, we only sum over the path segments where the atoms are in the excited state.

Explicit expressions for the Doppler, Shapiro, and Einstein phase shifts can be extracted for particular configurations. In what follows, we provide analytical phase shift expressions for two popular single-photon gradiometer designs: the Mach--Zehnder (MZ) and the large-momentum-transfer (LMT) gradiometers. For convenience, we will describe both experiments in a frame where the lasers are at rest in the absence of a metric perturbation.\footnote{One could also choose to describe the experiment in the mid-point trajectory frame~\cite{Overstreet:2020ftt}.}
By convention, the right arm receives the initial momentum deposition.

\subsubsection{Mach-Zehnder gradiometer}\label{sec:MZ}

\begin{figure*}[t!]
    \includegraphics[width=0.95\textwidth]{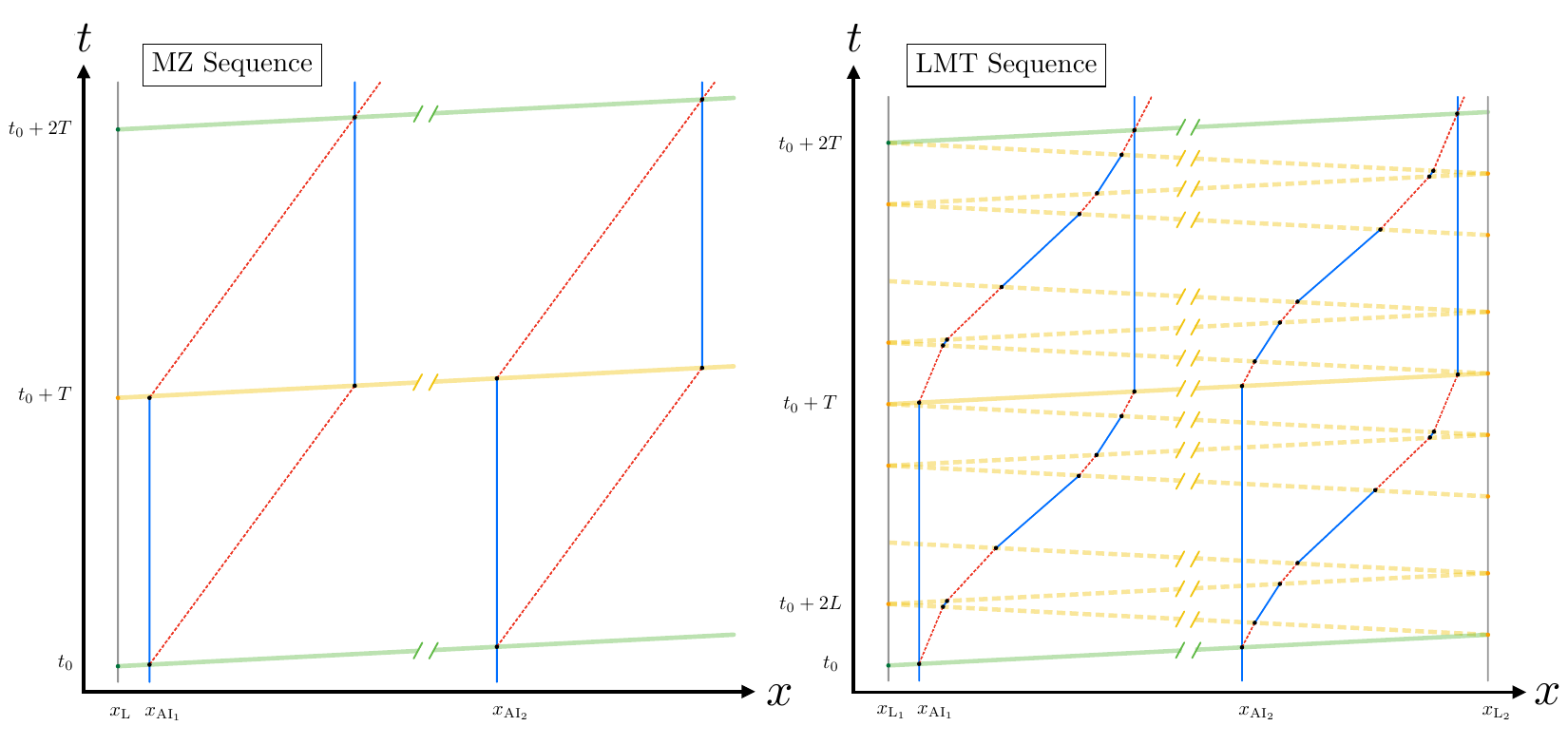} 
	\caption{Spacetime diagram of a gradiometer with Mach--Zehnder pulse sequence (left) and $n=4$ LMT pulse sequence (right). The laser worldlines are shown with grey solid lines, while the pulses they emit are shown in green for beam-splitter ($\pi/2$) pulses or in yellow for $\pi$ pulses; $\pi$ pulses that affect one (two) arm(s) of the AI are dashed (solid). Atoms in the ground state are shown with blue worldlines, while those in the excited state are shown in red.  The interaction with the atoms (marked with dots) of the $\pi/2$ pulses cause half of the atoms (corresponding to one arm or the other) to flip from ground to excited (or vice versa), while the $\pi$ pulses cause all the atoms to flip (from ground to excited or vice versa).  Note the LMT sequence features a total of $4 n - 1$ laser pulses. }\label{fig:sequence}
\end{figure*}

A MZ gradiometer employs a ``$\pi/2-\pi-\pi/2$" sequence. %
At time $t_0$,  which is also the time at which the atoms are released from their initial positions, a $\pi/2$-pulse is emitted from the laser to create an equal superposition of ground and excited state in both interferometers. As a result of atom-light interactions, the excited state wavepacket recoils with external momentum $k_\mathrm{eff}=\omega_a$.\footnote{Here, we assume that the laser is on resonance with the atom's transition frequency.} %
At time $t_0+T$, the laser emits a $\pi$-pulse, which swaps the atomic states in the arms of each AI, and reverts the relative momentum between the two wavepackets in each AI. The two arms of the interferometer are redirected before the application of a final beamsplitter pulse, which is emitted by the laser at time $t_0+2T$ and splits each arm into an equiprobable superposition of ground and excited states. This is followed by the interference measurement. 
The corresponding spacetime diagram is schematically depicted in the left panel of Fig.~\ref{fig:sequence}. 

In this configuration, the atom spends half of the time in the excited state. %
Denoting the locations of the two AIs as $x_{\mathrm{AI}_1}$ and $x_{\mathrm{AI}_2}=x_{\mathrm{AI}_1}+L$ and neglecting the distance between the laser source and the first AI, the gradiometer time delays for pairs of excited state paths initiated by a pulse emitted at unperturbed coordinate time $t$ can be written as
\begin{widetext}
\begin{equation} \label{eq:grad-time delays-MZ}
\begin{aligned}
\Delta t_{\mathrm{grad},\mathcal{E}}^\mathrm{MZ} (t) & = -\frac{1}{2}\left [\int_{t}^{t+T}  h_{00} (t',x_{\mathrm{AI}_1}) dt' - \int_{t +L }^{t + T + L}h_{00}(t',x_{\mathrm{AI}_2}) dt' \right] \, , \\ 
\Delta t_{\mathrm{grad},\mathcal{D}}^\mathrm{MZ}(t) & = n_i \Big [ \delta x^i(t+T,x_{\mathrm{AI}_1})- \delta x^i(t,x_{\mathrm{AI}_1}) +\delta x^i(t+L,x_{\mathrm{AI}_2})  -  \delta x^i(t+T+L,x_{\mathrm{AI}_2}) \Big ] \, , \\
\Delta t_{\mathrm{grad},\mathcal{S}}^\mathrm{MZ}(t) &= 
\int_{x_{\mathrm{AI}_1}}^{x_{\mathrm{AI}_2}}  \mathcal{H}_{+} (t(x'),x')dx' - \int_{x_{\mathrm{AI}_1}}^{x_{\mathrm{AI}_2}} \mathcal{H}_{+} (t(x')+T,x')dx' \, ,
\end{aligned}
\end{equation}
where $\delta x^i$ is given in Eq.~\eqref{eq:delta_x}, and $t(x')\equiv t+x'-x_{\mathrm{AI}_1}$ is the coordinate time parameterizing the initiating photon pulse.
Using Eqs.~\eqref{eq:grad-form}--\eqref{eq:grad-time delays-MZ}, the Einstein, Doppler and Shapiro gradiometer phase shifts for an experiment initiated at time $t_0$ may be compactly rewritten as
\begin{equation}\label{eq:grad-phase-MZ}
\begin{aligned}
\Delta \phi_{\mathrm{grad},\mathcal{E,D,S}}^\mathrm{MZ} (t_0) & = 
 \omega_a \Big( \Delta t_{\mathrm{grad},\mathcal{E,D,S}}^\mathrm{MZ} (t_0) - \Delta t_{\mathrm{grad},\mathcal{E,D,S}}^\mathrm{MZ} (t_0 + T ) \Big) \, .
\end{aligned}
\end{equation}

Equivalently, the three contributions may be expressed in the frequency domain. 
As we show in Appendix~\ref{app:MZ-config}, the Fourier transforms\footnote{Throughout the paper, we use the Fourier transform convention commonly adopted by the gravitational wave community: $\widetilde{x}(\omega)=\int_{-\infty}^{\infty}dt\,x(t)e^{-\mi\omega t}$ and $x(t)=(2\pi)^{-1}\int_{-\infty}^{\infty}d\omega \,\widetilde{x}(\omega)e^{\mi\omega t}$, where $\omega\equiv 2\pi f$~\cite{Moore:2014lga}.} of Eq.~\eqref{eq:grad-phase-MZ} can be compactly expressed as 
\begin{equation} \label{eq:grad-phase-MZ-FT}
\begin{aligned}
    \Delta \widetilde{\phi}^\mathrm{MZ}_{\mathrm{grad},\mathcal{E}}(\omega) %
    & =  \omega_a \, T^2 \, \omega^2\,
    K_{\rm MZ}   (\omega) \left [- \frac{1}{2\mi \omega} \widetilde{h}_{00} (\omega,x_{\mathrm{AI}}) e^{\mi \omega (x_{\mathrm{AI}} - x_{\mathrm{AI}_1})} \right ]_{x_{\mathrm{AI}_2}}^{x_{\mathrm{AI}_1}} \, , \\
    \Delta \widetilde{\phi}^\mathrm{MZ}_{\mathrm{grad},\mathcal{D}}(\omega) %
    & = \omega_a \, T^2 \, \omega^2\,
    K_{\rm MZ}   (\omega) \left [ n_i \delta \widetilde{x^i}(\omega, x_{\mathrm{AI}})e^{\mi \omega (x_{\mathrm{AI}} - x_{\mathrm{AI}_1})}\right]_{x_{\mathrm{AI}_2}}^{x_{\mathrm{AI}_1}} \, , \\
    \Delta \widetilde{\phi}^\mathrm{MZ}_{\mathrm{grad},\mathcal{S}}(\omega) %
    & = \omega_a \, T^2 \, \omega^2\,
    K_{\rm MZ}   (\omega) 
    \left[  
    \Delta \widetilde{\mathcal{T}}_\mathcal{S}^+ (\omega, x_{\mathrm{AI}_1}, x_{\mathrm{AI}}) \right ]_{x_{\mathrm{AI}_2}}^{x_{\mathrm{AI}_1}} \, ,
\end{aligned}
\end{equation}
\end{widetext}
where $\omega$ is the angular frequency and $\Delta \widetilde{\mathcal{T}}_\mathcal{S}^+$ is defined according to Eq.~\eqref{eq:TScurl}. For convenience, we define the response kernel for the MZ pulse sequence as
\begin{equation}\label{eq:MZ-response}
    \begin{split}
 K_{\rm MZ}   (\omega) =
 e^{\mi\omega T} \sinc^2\left(\frac{\omega T}{2}\right) \, ,
    \end{split}
\end{equation}
which is suppressed at angular frequencies ${\omega \gtrsim 1/2T}$ and asymptotically becomes a pure phase factor at low frequencies, as expected. 
Importantly, Eqs.~\eqref{eq:grad-phase-MZ-FT}--\eqref{eq:MZ-response} can be directly used to calculate the power spectrum density of the total AG phase shift in a MZ gradiometer induced by an arbitrary metric perturbation. %

\subsubsection{Large-momentum-transfer gradiometer}\label{sec:LMT}

The LMT sequence is a type of gradiometer that employs multiple ``kicks" to increase an AI's spacetime area, and therefore an AI's sensitivity. In this configuration, the ``beam-splitter sequence" and ``mirror sequence" consist of alternating laser pulses that are emitted from laser sources located at opposite ends of the baseline. Using the convention of Refs.~\cite{Graham:2016plp,Badurina:2022ngn}, a typical LMT sequence consists of $4n-1$ pulses, where $n$ is even. The first ``beam-splitter sequence" starts with a $\pi/2$-pulse emitted at $t_0$ by the laser source closest to the first AI; this is immediately followed by $n-1$ consecutive $\pi$-pulses which are emitted alternatingly from each laser and only interact with the right arm of each AI, thus incrementally depositing momentum onto the atoms. This is achieved by tuning the laser's frequency to the Doppler-shifted narrow transition frequency of the target atomic wavepacket. After a time $\sim n L$, both paths are in the ground state with relative velocity $n\omega_a/m_o$. The ``mirror sequence" 
consists of $2n-1$ consecutive $\pi$-pulses, the first of which is emitted at time $t_0+T-(n-1)L$ by the laser closest to the second AI; the first set of $n-1$ pulses interact with the right arm of each AI; the $\pi$-pulse emitted at time $t_0+T$ from the laser closest to the first AI interacts with \textit{both arms} of each AI; the remaining $n-1$ pulses interact with the left arm of each AI, thus bringing the relative velocity between the right and left arms to $-n\omega_a/m_o$. The final ``beam-splitter sequence" starts at time $t_0+2T-(n-1)L$ and consists of another $n-1$ $\pi$-pulses that bring the paths together. This is followed by a final $\pi/2$-pulse at time $t_0+2T$, which recombines the paths for measurements. For concreteness, we depict the associated spacetime diagram for $n=4$ in the right panel of Fig.~\ref{fig:sequence}.

Unlike the MZ sequence, an atom in this configuration is predominantly in the ground state. Moreover, the two AIs in the AG are not related by a simple spacetime translation. Therefore, the Einstein, Doppler and Shapiro phase shift terms associated with a particular excited state path segment depend on two factors: whether the path segment is initiated by a photon emitted by the laser located at $x_{\mathrm{L}_1}$ or $x_{\mathrm{L}_2}$, and whether the path segment is along the left or right arm of a given AI. In these configurations, the time difference between the emissions of the two common laser pulses defining a particular excited state path segment is $L$.\footnote{We assume that the second laser pulse is emitted immediately after the first pulse, which initiates an excited state path segment, and reaches the position of the second laser.} 
As shown in Fig.~\ref{fig:sequence}, the excited state segments can be grouped into ``beam-splitter" and ``mirror" sequences. Elements of these sequences can be classified as ``$>$" or ``$<$" segments, based on whether the excited state segments are initiated by a pulse from $x_{\mathrm{L}_1}$ or $x_{\mathrm{L}_2}$, respectively.
The spacetime diagram also illustrates that an excited state pulse along an arm of an AI is followed by 
$n/2-1$ excited state path segments, all of which are initiated by pulses from the same source. 

Similar to the MZ configuration, each ``arm" in an LMT configuration that interacts with the laser pulses alternates between excited and ground state with each atom-light interaction. For each ``beam-splitter" and ``mirror" sequence, the path segments are labeled by $k$, which runs from $0$ to $n-1$, with even and odd values of $k$ labeling excited and ground state path segments, respectively. This enables us to assign the coordinate time label $t_k(t,x_\mathrm{AI})$ to each atom-light interaction point for atoms at unperturbed position $x_\mathrm{AI}$ that interact with the $k$-th laser pulse in a given ``beam-splitter" or ``mirror" sequence beginning at time $t$, i.e. 
\begin{equation}\label{eq:t_k}
	t_k (t, x) = t+kL+{(\pm)}_{(k)}\left (x-x_{\mathrm{L}}^{(k)}\right ) \, ,
\end{equation}
where $x_\mathrm{L}^{(k)}$ is the location of the laser source that emits the $k$-th pulse, and ${(\pm)}_{(k)}$ corresponds to the direction of the $k$-th laser pulse. An ``outgoing" pulse is defined to be parallel to the AG baseline $\unit{n}$ and takes the sign ``$+$", while an ``incoming" pulse is anti-parallel to the baseline and takes the sign ``$-$". In an LMT sequence, a pair of laser pulses that define a path segment always consists of two consecutive pulses from opposite ends of the baseline. There are two possible combinations of laser pulse pairs: ``$>$" consists of an outgoing-incoming pair, and ``$<$" consists of an incoming-outgoing pair. We define the gradiometer time delays for a path segment initiated by a pulse from $x_{\mathrm{L}_1}$ (i.e. a ``$>$" segment) to be $\Delta t^{(k)}_{\text{grad},\mathcal{D,S,E}} \big|_{\mathrm{L}_1}$. In this case, pulse $k$ is emitted from $x_{\mathrm{L}}^{(k)}=x_{\mathrm{L}_1}$ and is outgoing (i.e. ${(\pm)}_{(k)}=+1$), while pulse $k+1$ is emitted from $x_{\mathrm{L}}^{(k+1)}=x_{\mathrm{L}_2}$ and is incoming (i.e. ${(\pm)}_{(k+1)}=-1$). Similarly, the gradiometer time delays for a path segment initiated by a pulse from $x_{\mathrm{L}_2}$ (i.e. a ``$<$" segment) is written as $\Delta t^{(k)}_{\text{grad},\mathcal{D,S,E}} \big|_{\mathrm{L}_2}$. In this case, pulse $k$ is emitted from $x_{\mathrm{L}}^{(k)}=x_{\mathrm{L}_2}$ and is incoming (i.e. ${(\pm)}_{(k)}=-1$), while pulse $k+1$ is emitted from $x_{\mathrm{L}}^{(k+1)}=x_{\mathrm{L}_1}$ and is outgoing (i.e. ${(\pm)}_{(k+1)}=+1$). With this convention, the gradiometer time delays can be written as:
\begin{widetext}
\begin{equation}\label{eq:grad-time delays-LMT-L1L2}
\begin{aligned}
\left. \Delta t_{\mathrm{grad},\mathcal{E}}^{(k)} (t)\right|_{\mathrm{L}_1,\mathrm{L}_2} 
& = -\frac{1}{2}\Bigg [\int_{t_k(t,x_\mathrm{AI})}^{t_{k+1}(t,x_\mathrm{AI})}  h_{00} (t',x_{\mathrm{AI}}) dt' \Bigg] \Bigg |^{x_{\mathrm{AI}_1}} _{x_{\mathrm{AI}_2}}\,  , \\
\left. \Delta t_{\mathrm{grad},\mathcal{D}}^{(k)} (t) \right|_{\mathrm{L}_1,\mathrm{L}_2} & = n_i \Big [{(\pm)}_{(k+1)}\delta x^i(t_{k+1}(t,x_\mathrm{AI}),x_{\mathrm{AI}}) - {(\pm)}_{(k)}\delta x^i(t_k(t,x_\mathrm{AI}),x_{\mathrm{AI}}) \Big ] \Bigg |^{x_{\mathrm{AI}_1}} _{x_{\mathrm{AI}_2}}\, , \\
\left. \Delta t_{\mathrm{grad},\mathcal{S}}^{(k)} (t)\right|_{\mathrm{L}_1,\mathrm{L}_2} 
& = \left[\Delta\mathcal{T}^{{(\pm)}_{(k+1)}}_\mathcal{S}\left (t+(k+1)L, x^{(k+1)}_\mathrm{L}, x_{\mathrm{AI}}\right) -\Delta\mathcal{T}^{{(\pm)}_{(k)}}_\mathcal{S}\left (t+kL, x^{(k)}_\mathrm{L}, x_{\mathrm{AI}}\right)\right] \Bigg |^{x_{\mathrm{AI}_1}} _{x_{\mathrm{AI}_2}} \, .
\end{aligned} 
\end{equation}
Notice that, differently from the Einstein time delay, the Doppler and Shapiro time delays originate from photon propagation and are therefore sensitive to the direction of the laser pulses. See App.~\ref{app:LMT-config} for a detailed derivation. 

Let us now consider the gradiometer phase shift for an experiment initiated at time $t_0$. In this case, sets of ``$>$" excited states segments are initiated at times $t_0$ and $t_0+T$, while sets of ``$<$" excited states segments are initiated at times $t_0+T-(n-1)L$ and $t_0+2T-(n-1)L$. Using Eqs.~\eqref{eq:t_k}--\eqref{eq:grad-time delays-LMT-L1L2}, the Einstein, Doppler and Shapiro contributions may be compactly expressed as 
\begin{equation}\label{eq:LMT_final}
\begin{aligned}
\Delta \phi^\mathrm{LMT}_{\mathrm{grad},\mathcal{E}, \mathcal{D}, \mathcal{S}}(t_0) &= \omega_a \sum_{\substack{k=0 \\ \text{even}}}^{n-2} \Bigg (\Delta t_{\mathrm{grad},\mathcal{E}, \mathcal{D}, \mathcal{S}}^{(k)} (t_0) \bigg |_{\mathrm{L}_1} + \Delta t_{\mathrm{grad},\mathcal{E},\mathcal{D}, \mathcal{S}}^{(k)} (t_0+T-(n-1)L) \bigg |_{\mathrm{L}_2} \\
& \qquad \qquad \quad - \Delta t_{\mathrm{grad},\mathcal{E},\mathcal{D}, \mathcal{S}}^{(k)} (t_0+T) \bigg |_{\mathrm{L}_1} - \Delta t_{\mathrm{grad},\mathcal{E}, \mathcal{D}, \mathcal{S}}^{(k)} (t_0+2T-(n-1)L) \bigg |_{\mathrm{L}_2}\Bigg ) \, ,
\end{aligned}
\end{equation}
which can be summed to produce the overall gauge-invariant LMT observable.
 
Interestingly, the summation over the excited state segments can be performed explicitly in the frequency domain, leading to a closed-form expression (see~Appendix~\ref{app:LMT-config} for a derivation). Indeed, the three contributions to the AG phase shift may be compactly rewritten as
\begin{equation} \label{eq:LMT_tot_F}
\begin{aligned}
\Delta \widetilde{\phi}^\mathrm{LMT}_{\mathrm{grad},\mathcal{E}} (\omega) & =  \frac{1}{2} k_{\rm eff} \, T^2 \,  \omega^2 \,K_{\rm MZ}   (\omega)\, 
 \left [ - \frac{1}{2 \mi \omega} \widetilde{h}_{00} (\omega,x_{\mathrm{AI}}) \left (    
 K_{\rm LMT}^+ (\omega)e^{\mi \omega (x_{\mathrm{AI}}-x_{\mathrm{L}_1})} -K_{\rm LMT}^- (\omega)  e^{\mi \omega (x_{\mathrm{L}_2}-x_{\mathrm{AI}})}\right )  
\right]_{x_{\mathrm{AI}_2}}^{x_{\mathrm{AI}_1}} \, , \\
\Delta \widetilde{\phi}^\mathrm{LMT}_{\mathrm{grad},\mathcal{D}} (\omega) & = \frac{1}{2} k_{\rm eff} \, T^2 \,  \omega^2 \,K_{\rm MZ}   (\omega)\, 
 \left [n_i \delta \widetilde{x}^i (\omega,x_{\mathrm{AI}}) \left (K_{\rm LMT}^+ (\omega)   e^{\mi \omega (x_{\mathrm{AI}}-x_{\mathrm{L}_1})} +  K_{\rm LMT}^- (\omega) e^{\mi \omega (x_{\mathrm{L}_2}-x_{\mathrm{AI}})}\right )
\right]_{x_{\mathrm{AI}_2}}^{x_{\mathrm{AI}_1}} \, , \\
\Delta \widetilde{\phi}^\mathrm{LMT}_{\mathrm{grad},\mathcal{S}} (\omega) & = \frac{1}{2} k_{\rm eff} \, T^2 \,  \omega^2 \,K_{\rm MZ}   (\omega)\, 
 \left [ K_{\rm LMT}^+ (\omega) \Delta \widetilde{\mathcal{T}}^+_{\mathcal{S}} (\omega,x_{\mathrm{L}_1},x_{\mathrm{AI}})  -  K_{\rm LMT}^- (\omega) \Delta \widetilde{\mathcal{T}}^-_{\mathcal{S}} 
(\omega, x_{\mathrm{L}_2},x_{\mathrm{AI}}  )
\right]_{x_{\mathrm{AI}_2}}^{x_{\mathrm{AI}_1}} \, .
\end{aligned}
\end{equation}
Here, $k_{\rm eff} = n \omega_a$. The structure of the individual phase shift contributions resembles that of the MZ case, as is evidenced by the presence of the kernel 
$K_\mathrm{MZ}(\omega)$ (\textit{cf}.~Eq.~\eqref{eq:grad-phase-MZ-FT}). However, unlike the MZ case, %
the individual gradiometer phase shift contributions contain additional frequency-dependent response kernels, 
which depend on the laser beams' direction of propagation and on the number of LMT kicks $n$. These LMT--specific response kernels 
are defined in terms of experimental parameters as
\begin{equation} \label{eq:KLMTall}
\begin{split}
K^{\pm}_{\rm LMT}(\omega)
= \frac{\sinc \left (\dfrac{n\omega L}{2} \right)}{\sinc \left (\dfrac{\omega T}{2} \right ) \sinc ( \omega L)}
\begin{cases}
& \sinc \left (\dfrac{\omega (T-(n-2) L)}{2}\right ) \left(1-\dfrac{(n-2) L}{T}\right) 
\quad \text{for outgoing photons~} (+)  \,  ,
\\
& \sinc \left (\dfrac{\omega (T-nL)}{2} \right )\left(1-\dfrac{nL}{T}\right) 
\quad \text{for incoming photons~} (-) \,. 
\\
\end{cases}
\end{split}
\end{equation}
\end{widetext}
Notably, these 
functions give rise to an additional parametric suppression of the signal for frequencies $\omega \gtrsim 1/nL$. Furthermore, the combination of the LMT kernels in 
Eq.~\eqref{eq:LMT_tot_F} further suppresses the Einstein gradiometer phase shift relative to the Doppler and Shapiro contributions. One can best observe this in the low frequency limit and for $n \gg 1$: in this regime, $K^{+}_{\rm LMT} + K^{-}_{\rm LMT}\approx~2~K^{+}_{\rm LMT}$ and $K^{+}_{\rm LMT} - K^{+}_{\rm LMT} \approx 2L/T$. Hence, $\Delta \widetilde{\phi}^\mathrm{LMT}_{\mathrm{grad},\mathcal{E}} (\omega)$ receives an additional $L/T$ suppression.

\section{Applications}\label{sec:applications}

Equipped with the formalism introduced in the previous section, we now compute the leading order phase shift in both MZ and LMT configurations for two well-motivated examples: a transient gravitational wave (GW) and a slow-varying weak Newtonian potential. 

\subsection{Gravitational waves}\label{sec:GW}
The first application that we consider in this work is the signal induced by a transient GW. %
By working in different frames and with different single-photon gradiometer configurations, we demonstrate that our result: \textit{(i)} is invariant under general coordinate transformations and \textit{(ii)} accurately reproduces results from the literature. Here, we note that an analogous calculation in these two gauges, within the context of a laser interferometer, is well documented in the literature (see, e.g., Ref.~\cite{Maggiore:2007ulw}). 

\subsubsection{Transverse-traceless gauge}

Firstly we compute the phase shift in the transverse-traceless (TT) gauge. Assuming that the GW propagates in the $z$ direction, in TT-gauge the metric tensor perturbation takes the form
\begin{equation}
h_{ij}^\mathrm{TT}(t, z) = \begin{pmatrix}
h_+ & h_\times & 0 \\
h_\times & -h_+ & 0 \\
0 & 0 & 0
\end{pmatrix}_{ij}
\cos \left (\omega_\mathrm{g} (t-z) +\theta \right ) \, .
\end{equation}
Here, $i,j \in \{1,2,3\}$, $h_\times$ and $h_+$ are the amplitudes of the two polarizations, $\omega_\mathrm{g}$ is the GW angular frequency, and $\theta$ is the phase of the GW at $(t,z)=(0,0)$ ~\cite{Maggiore:2007ulw}. Without loss of generality, we set the gradiometer at $z=0$ and the baseline in the $x$-direction. To make contact with the literature, we work in the very long-baseline limit, i.e. $L = x_{\mathrm{L}_2}-x_{\mathrm{L}_1} \approx x_{\mathrm{AI}_2}-x_{\mathrm{AI}_1} $. In this gauge, the Einstein phase shift, which only depends on $h_{00}$, vanishes (cf. Eq.~\eqref{eq:E0_1}). The Doppler phase shift, which depends on both $h_{i0}$ and $h_{00}$ through the atom's geodesic equation, also vanishes (cf. Eqs.~\eqref{eq:deltav},~\eqref{eq:delta_x} and \eqref{eq:Doppler_schematic}), regardless of the gradiometer configuration.  Hence, in this gauge, the phase shift can only depend on the Shapiro time delay. 

Let us first focus on the simpler MZ gradiometer configuration.
Making use of Eqs.~\eqref{eq:grad-time delays-MZ}--\eqref{eq:grad-phase-MZ}, %
the leading order phase shift in this gauge takes the simple form %
\begin{equation}\label{eq:MZ_GW_TT_1}
\begin{aligned}
\Delta  \phi_\mathrm{grad}^{\mathrm{MZ}}(t_0) %
& = -2 h_+ L \, \omega_a  \sinc \left(\frac{\omega_\mathrm{g} L}{2}\right) \sin^2 \left(\frac{\omega_\mathrm{g} T}{2}\right) 
\cos \left (\omega_\mathrm{g} t_0 +\omega_\mathrm{g} \left(T+\frac{L}{2}\right)+ \theta \right) \, ,
\end{aligned}
\end{equation} 
whose amplitude can be rewritten as
\begin{equation}\label{eq:MZ_GW_TT}
\begin{aligned}
\left |\Delta \phi_\mathrm{grad}^{\mathrm{MZ}}\right | = 2 h_+ L \, \omega_a  \left |\sinc \left(\frac{\omega_\mathrm{g} L}{2}\right) \right |\sin^2 \left(\frac{\omega_\mathrm{g} T}{2}\right)\, .
\end{aligned}
\end{equation}

Similarly,
using Eqs.~\eqref{eq:t_k}--\eqref{eq:LMT_final}, the phase shift induced by a transient GW in an LMT gradiometer takes the form %
\begin{equation}\label{eq:LMT_GW_TT_1}
\begin{aligned}
\Delta \phi_{\mathrm{grad}}^{\mathrm{LMT}}(t_0) =& -2 h_+ L 
k_{\rm eff}
\sinc \left ( \frac{n\omega_\mathrm{g} L}{2} \right )\sin \left ( \frac{\omega_\mathrm{g} T}{2} \right ) 
\sin \left ( \frac{\omega_\mathrm{g}\left(T-(n-1)L\right)}{2}\right ) 
\cos \left (\omega_\mathrm{g} t_0 +\omega_\mathrm{g} \left(T+\frac{L}{2}\right)+ \theta \right) \, ,
\end{aligned}
\end{equation}
where $k_{\rm eff} = n \, \omega_a $. The amplitude of the signal can be rewritten as
\begin{equation}\label{eq:LMT_GW_TT}
\begin{aligned}
\left |\Delta \phi_{\mathrm{grad}}^{\mathrm{LMT}} \right | &= 2 h_+ L k_{\rm eff} \left | \sinc \left ( \frac{n\omega_\mathrm{g} L}{2} \right )\sin \left ( \frac{\omega_\mathrm{g} T}{2} \right ) 
\sin \left ( \frac{\omega_\mathrm{g}\left(T-(n-1)L\right)}{2}\right )  \right | \, 
.
\end{aligned}
\end{equation}

Notably, Eqs.~\eqref{eq:MZ_GW_TT} and \eqref{eq:LMT_GW_TT} agree with previous results from the literature~\cite{Graham:2012sy,Graham:2016plp}. Furthermore, our result highlights the origin of the effect in TT-gauge: since test masses in free-fall are unaffected by a transient GW in TT-gauge, the effect of interest appears solely through the delay in the arrival time of photons between the two AIs. 

This analysis can also be repeated in the frequency domain. Using the master equation for the MZ configuration (\textit{cf}.~Eq.~\eqref{eq:grad-phase-MZ-FT}), the Fourier transform of the Shapiro time delay (\textit{cf}.~Eq.~\eqref{eq:TScurl}) and the time-dependence of the GW's phase, we find that the MZ phase shift takes the form %
\begin{equation}\label{eq:MZ-FT-GW}
\begin{aligned}
\Delta \widetilde{\phi}_{\mathrm{grad}}^{\mathrm{MZ}}(\omega) &= -\omega_a T^2  \omega^2 K_{\mathrm{MZ}}(\omega) \Delta \widetilde{\mathcal{T}}^+_{\mathcal{S}}(\omega,0,L)  \\ &= -\frac{\pi}{2}\omega_a h_+L\omega^2T^2  K_{\mathrm{MZ}}(\omega)  \sinc \left (\frac{\omega L}{2}\right) 
e^{\mi\theta}e^{\mi \omega L/2} \delta (\omega-\omega_\mathrm{g})  \, ,
\end{aligned}
\end{equation}
where we restricted our attention to positive frequencies. Similarly, using the master equation for the LMT configuration (\textit{cf}.~Eq.~\eqref{eq:LMT_tot_F}) 
the gradiometer phase shifts in the frequency domain can be expressed as
\begin{widetext}
\begin{equation}\label{eq:LMT-FT-GW}
\begin{aligned}
\Delta \widetilde{\phi}_{\mathrm{grad}}^{\mathrm{LMT}}(\omega) &= -\frac{1}{2} k_{\rm eff} \, T^2 \,  \omega^2 \,K_{\rm MZ}   (\omega)  
 \left ( K_{\rm LMT}^+ (\omega) \Delta \widetilde{\mathcal{T}}^+_{\mathcal{S}} (\omega,0,L) +K_{\rm LMT}^- (\omega) \Delta \widetilde{\mathcal{T}}^-_{\mathcal{S}} (\omega,L,0) \right ) \\
 & = -\frac{\pi}{4}k_{\mathrm{eff}}h_+L\omega^2T^2  \sinc \left (\frac{\omega L}{2}\right ) e^{\mi \theta}e^{\mi \omega L/2}K_{\rm MZ}(\omega)\left[K_{\rm LMT}^+(\omega)+K_{\rm LMT}^-(\omega)\right] \delta(\omega-\omega_\mathrm{g})\, .
\end{aligned}
\end{equation}
\end{widetext}
Importantly, after some algebra, one can show that Eqs.~\eqref{eq:MZ-FT-GW} and \eqref{eq:LMT-FT-GW} agree exactly with the Fourier transform of Eqs.~\eqref{eq:MZ_GW_TT_1} and \eqref{eq:LMT_GW_TT_1}, respectively, thus validating the form of the master equations in section~\ref{sec:framework_derivation:definite_pulses}.

\subsubsection{Proper detector frame}

To explicitly verify that our result is invariant under general coordinate transformations, we now repeat the GW calculation in the gradiometer's proper detector frame. The coordinate system in this frame is constructed along the worldline of a freely falling (i.e. non-accelerating) observer by extending spacelike vectors orthogonal to the observer’s four velocity~\cite{Rakhmanov:2014noa}. In this frame, the metric takes the form 
\begin{equation}
ds^2 = (-1+\widehat{h}_{00})dt^2+\widehat{h}_{0i} dt dx^i + (1+\widehat{h}_{ij})dx^i dx^j \, ,
\end{equation}
where the components of the metric perturbation tensor are defined as
\begin{equation}
\begin{aligned}
\widehat{h}_{00} & =-\sum_{r=0}^{\infty} \frac{2}{(r+2)!} \widehat{R}_{0 m 0 n, k_1 \cdots k_r} x^m x^n x^{k_1} \cdots x^{k_r} \, , \\
\widehat{h}_{{0} i} & =-\sum_{r=0}^{\infty} \frac{2(r+2)}{(r+3)!} \widehat{R}_{{0 m i n}, k_1 \cdots k_r} x^m x^n x^{k_1} \cdots x^{k_r} \, ,\\
\widehat{h}_{{i j}} & =-\sum_{r=0}^{\infty} \frac{2(r+1)}{(r+3)!} \widehat{R}_{i m j n, k_1 \cdots k_r} x^m x^n x^{k_1} \cdots x^{k_r} \, .
\end{aligned}
\end{equation}
Here, the comma preceding the indices $k_1 \cdots k_r$ denotes differentiation with respect to these coordinates and the hat above any function implies that this function is evaluated along the observer's worldline $x_\mathrm{obs}$ e.g.,
\begin{equation}
\left.\hat{R}_{\mu \nu \rho \sigma, k_1 \ldots k_r} \equiv \frac{\partial^r R_{\mu \nu \rho \sigma}}{\partial x^{k_1} \ldots \partial x^{k_r}}\right|_{x=x_\mathrm{obs}} \, ,
\end{equation}
with ${R}_{\mu \nu \rho \sigma} = (\partial_{\nu}\partial_\rho h_{\mu\sigma}+\partial_{\mu}\partial_\sigma h_{\nu\rho}-\partial_{\mu}\partial_\rho h_{\nu\sigma}-\partial_\nu \partial_\sigma h_{\mu \rho})/2$ the Riemann tensor in linearized gravity~\cite{Marzlin:1994ia}. In this frame, $x^{i}$ is the spacelike separation between the observer's worldline and an arbitrary spacetime point at fixed time; setting $x^{i} = 0$ we recover the metric of flat spacetime. %
Importantly, the Riemann tensor is invariant (element-wise) under gauge transformations to leading order in the metric perturbation. Therefore, without loss of generality, we may evaluate it in TT-gauge. Since the gradiometer probes test mass dynamics along a single spatial component, and we assumed that the GW travels in the $z$-direction with the atoms at $z=0$ and moving only along the $x$-direction (i.e. $i = 1$), one can easily show that the metric tensor perturbation in this frame takes the especially simple form~\cite{Berlin:2021txa}
\begin{equation}\label{eq:PD_metric}
\begin{aligned}
\widehat{h}_{\mu\nu} &= \widehat{h}_{00}^\mathrm{TT} \, , \\
\widehat{h}_{00}^\mathrm{TT} &= -(x^{i})^2 \widehat{R}^{\mathrm{TT}}_{0i0i} = \frac{x^2}{2} \partial_0^2 h_{11}^{\mathrm{TT}} \\ & = -\frac{\omega_\mathrm{g}^2x^2}{2}h_+\cos(\omega_\mathrm{g}(t-z)+\theta) \, .
\end{aligned}
\end{equation}
From the results of section~\ref{sec:framework_derivation:definite_pulses}, we observe that in this frame all contributions to the gradiometer phase shift are non-zero.

Using Eq.~\eqref{eq:PD_metric}, we compute the phase shift for the MZ and LMT pulse sequences in the proper detector frame. %
Using our master equations and working in the limit $x_{\mathrm{AI}_2}-x_{\mathrm{AI}_1} \approx L$, the Doppler, Shapiro and Einstein phase shift contributions to the MZ observable take the form %
\begin{widetext}
\begin{equation}\label{eq:MZ_GW_PD_EDS}
\begin{aligned}
\Delta \phi^\mathrm{MZ}_{\mathrm{grad}, \mathcal{D}}(t_0) = &-2 h_+ L \omega_a  \sin^2 \left (\frac{\omega_\mathrm{g} T}{2}\right ) \cos \left (\omega_\mathrm{g} t_0+\omega_\mathrm{g} (T+L) + \theta \right ) \, , \\
\Delta \phi^\mathrm{MZ}_{\mathrm{grad}, \mathcal{S}}(t_0) = &+2 h_+ L \omega_a  \sin^2 \left (\frac{\omega_\mathrm{g} T}{2}\right ) \cos \left (\omega_\mathrm{g} t_0 + \omega_\mathrm{g} (T+L) + \theta \right ) \\
&-2 h_+ L \, \omega_a  \sinc \left(\frac{\omega_\mathrm{g} L}{2}\right) \sin^2 \left(\frac{\omega_\mathrm{g} T}{2}\right)\cos \left (\omega_\mathrm{g} t_0+\omega_\mathrm{g}  \left(T+  \frac{L}{2}\right)+ \theta \right)  \\ &+ h_+ L^2 \omega_a \omega_\mathrm{g} \sin^2 \left (\frac{\omega_\mathrm{g} T}{2}\right ) \sin \left (\omega_\mathrm{g} t_0+\omega_\mathrm{g}(T+L) + \theta \right ) \, ,\\
\Delta \phi^\mathrm{MZ}_{\mathrm{grad}, \mathcal{E}}(t_0) = &- h_+ L^2 \omega_a \omega_\mathrm{g} \sin^2 \left (\frac{\omega_\mathrm{g} T}{2}\right ) \sin \left (\omega_\mathrm{g} t_0+ \omega_\mathrm{g}(T+L) + \theta \right ) \, .
\end{aligned}
\end{equation}
Adding the three contributions, we exactly recover Eq.~\eqref{eq:MZ_GW_TT_1}, as desired. To gain a better understanding of the effects induced by a transient gravitational wave on the AG, it is advantageous to take the limit $\omega_\mathrm{g} L \ll 1$ and isolate the leading order phase shifts in Eq.~\eqref{eq:MZ_GW_PD_EDS}. Explicitly, 
\begin{equation}\label{eq:MZ_GW_PD_EDS_wL}
\begin{aligned}
\Delta \phi^\mathrm{MZ}_{\mathrm{grad}, \mathcal{D}}(t_0) &= -2 h_+ L \omega_a  \sin^2 \left (\frac{\omega_\mathrm{g} T}{2}\right ) \cos \left (\omega_\mathrm{g} t_0+\omega_\mathrm{g} T + \theta \right ) + \mathcal{O}(\omega_\mathrm{g} L)\, , \\
\Delta \phi^\mathrm{MZ}_{\mathrm{grad}, \mathcal{S}}(t_0)  &=\frac{1}{3} h_+\, L^3\omega_a\omega_\mathrm{g}^2 \sin^2 \left (\frac{\omega_\mathrm{g} T}{2}\right ) \cos (\omega_\mathrm{g} t_0+\omega_\mathrm{g} T+ \theta ) + \mathcal{O}(\omega_\mathrm{g}^3 L^3)\, ,\\
\Delta \phi^\mathrm{MZ}_{\mathrm{grad}, \mathcal{E}}(t_0)  &= - h_+ L^2 \omega_a \omega_\mathrm{g} \sin^2 \left (\frac{\omega_\mathrm{g} T}{2}\right ) \sin \left (\omega_\mathrm{g} t_0+\omega_\mathrm{g} T + \theta \right ) + \mathcal{O}(\omega_\mathrm{g}^2 L^2) \, .
\end{aligned}
\end{equation}
\end{widetext}
From Eq.~\eqref{eq:MZ_GW_PD_EDS_wL}, the hierarchy of contributions is evident: in this frame, the Doppler phase shift dominates, while the Einstein and Shapiro phase shift are subdominant, with the latter being further suppressed with respect to the leading order Einstein phase shift. This can be understood as follows. In the PD frame, the effect of a GW can be described in terms of Newtonian forces. Indeed, $\partial_0^2 x^i = h_{ij}^{\mathrm{TT}}x^j/2$, which follows from the geodesic deviation equation~\cite{Maggiore:2007ulw}; in turn this implies that $\hat{h}_{00} = x^i a_i$, where $a^i \equiv \partial_0^2 x^i$ is the acceleration induced by the gravitational wave on photons and atoms. 
In NR computational frameworks (e.g.~\cite{Storey:1994oka}), the leading order phase shift due to forces acting on atoms scales as $k_\mathrm{eff} a T^2$. Hence, the Doppler phase shift contribution, which precisely accounts for the $\mathcal{O}(h)$ acceleration of the atoms in the AG, recovers this term at leading order.\footnote{In the limit $\omega T, \omega L \ll 1$, the amplitude of the leading order contribution to the Doppler phase shift is $h_+ L \omega_a \omega^2 T^2/2$. By making the identification $a \equiv h_+ L \omega^2/2$, we recover the desired result. Note that this also agrees with Eq.~\eqref{eq:grad-phase-MZ-FT} in the limit $\omega_\mathrm{g} \rightarrow 0$.} The Einstein and Shapiro contributions are pure GR effects, since the former depends on the rest mass correction of excited state path segments while the latter depends on the finite speed of light. It is therefore unsurprising that these effects are parametrically suppressed with respect to the leading order Doppler phase shift.

The calculation for the LMT pulse configuration follows analogously. Using the definitions in section~\ref{sec:framework_derivation}, the gradiometer phase shift contributions take the form %
\begin{widetext}
\begin{equation}\label{eq:LMT_GW_PD_EDS}
\begin{aligned}
\Delta \phi^\mathrm{LMT}_{\mathrm{grad}, \mathcal{D}}(t_0) &=   2\frac{h_+ L \omega_a}{\sin \left (\omega_\mathrm{g} L \right )} \sin\left (\frac{n\omega_\mathrm{g} L}{2} \right ) \sin \left ( \frac{\omega_\mathrm{g} T}{2} \right ) \\ & \qquad \times  \left \{ \sin \left (\omega_\mathrm{g} t_0+\omega_\mathrm{g}\left(\frac{n L + T}{2}\right) + \theta \right )  - 
    \cos(\omega_\mathrm{g} L) \sin \left (\omega_\mathrm{g} t_0+ \omega_\mathrm{g}\left(\frac{3 T - (n-2)L }{2}\right) + \theta \right ) \right \} \, , \\
\Delta \phi^\mathrm{LMT}_{\mathrm{grad}, \mathcal{S}}(t_0) &= -2 h_+ L k_{\rm eff} \sinc \left ( \frac{n\omega_\mathrm{g} L}{2} \right )\sin \left ( \frac{\omega_\mathrm{g} T}{2} \right )\sin \left ( \frac{\omega_\mathrm{g}\left(T-(n-1)L\right)}{2}\right ) \cos \left (\omega_\mathrm{g} t_0 +\omega_\mathrm{g} \left(T+\frac{L}{2}\right)+ \theta \right ) \, \\
&\qquad +h_+ L^2 \omega_\mathrm{g} \omega_a \sin \left (\frac{n\omega_\mathrm{g}  L}{2} \right ) \sin \left ( \frac{\omega_\mathrm{g} T}{2} \right ) \sin \left (\omega_\mathrm{g} t_0 + \omega_\mathrm{g}\left(\frac{ 3T-(n-2) L}{2}\right)
    + \theta \right )  \\ & \qquad \, - 2\frac{h_+ L \omega_a}{\sin \left (\omega_\mathrm{g} L \right )} \sin\left (\frac{n\omega_\mathrm{g} L}{2} \right ) \sin \left ( \frac{\omega_\mathrm{g} T}{2} \right )\\ & \times  \left \{ \sin \left (\omega_\mathrm{g} t_0+\omega_\mathrm{g}\left(\frac{n L + T}{2}\right) + \theta \right )  - 
    \cos(\omega_\mathrm{g} L) \sin \left (\omega_\mathrm{g} t_0+ \omega_\mathrm{g}\left(\frac{3 T - (n-2)L }{2}\right) + \theta \right ) \right \}  ,\\
\Delta \phi^\mathrm{LMT}_{\mathrm{grad}, \mathcal{E}}(t_0) &= -h_+ L^2 \omega_\mathrm{g} \omega_a \sin \left (\frac{n\omega_\mathrm{g}  L}{2} \right ) \sin \left ( \frac{\omega_\mathrm{g} T}{2} \right ) \sin \left (\omega_\mathrm{g} t_0 + \omega_\mathrm{g}\left(\frac{ 3T-(n-2) L}{2}\right)
    + \theta \right ) \, .
\end{aligned}
\end{equation}
As for the MZ case, the sum of the terms in Eq.~\eqref{eq:LMT_GW_PD_EDS} exactly recovers the TT-gauge result in Eq.~\eqref{eq:LMT_GW_TT_1}. In the long wavelength limit, the result exhibits interesting features. Explicitly, %
\begin{equation}\label{eq:LMT_GW_PD_EDS_wL}
\begin{aligned}
\Delta \phi^\mathrm{LMT}_{\mathrm{grad}, \mathcal{D}}(t_0) &= -2 h_+ L k_\mathrm{eff}  \sin^2 \left (\frac{\omega_\mathrm{g} T}{2}\right ) \cos \left (\omega_\mathrm{g} t_0+\omega_\mathrm{g} T + \theta \right ) + \mathcal{O}(\omega_\mathrm{g} L)\, , \\
\Delta \phi^\mathrm{LMT}_{\mathrm{grad}, \mathcal{S}}(t_0) &=\frac{1}{3} h_+\, L^3\omega_\mathrm{g}^2 k_\mathrm{eff} \sin^2 \left (\frac{\omega_\mathrm{g} T}{2}\right ) \cos (\omega_\mathrm{g} t_0+\omega_\mathrm{g} T+ \theta ) + \mathcal{O}(\omega_\mathrm{g}^3 L^3)\, ,\\
\Delta \phi^\mathrm{LMT}_{\mathrm{grad}, \mathcal{E}}(t_0) &= - \frac{1}{2} h_+ L^3 k_\mathrm{eff} \omega_\mathrm{g}^2 \sin \left (\frac{\omega_\mathrm{g} T}{2}\right ) \sin \left (\omega_\mathrm{g} t_0+\frac{3 \omega_\mathrm{g} T}{2} + \theta \right ) + \mathcal{O}(\omega_\mathrm{g}^3 L^3) \, .
\end{aligned}
\end{equation}
\end{widetext}
While the Shapiro and Doppler contributions are structurally equivalent to their respective contributions in Eq.~\eqref{eq:MZ_GW_PD_EDS_wL}, the leading-order LMT Einstein phase shift term is further suppressed by a factor of $L/T$ in the limit $\omega T \ll 1$, as anticipated at the end of section~\ref{sec:framework_derivation:definite_pulses} for a generic metric perturbation.\footnote{Interestingly, by relating the number of LMT kicks $n$ to the cavity finesse, the Doppler phase shift scales analogously to the leading order phase shift in a single-arm laser interferometer employing Fabry-Perot cavities~\cite{Maggiore:2007ulw}.}%

In summary, we explicitly showed that, for a transient GW, the phase shift obtained in TT-frame agrees with the observable computed in the PD frame. Furthermore, the results obtained using our simplified framework agree with previous work. This confirms the gauge-invariant nature of the gradiometer observable derived in section~\ref{sec:framework}.

\subsection{Slow-varying weak Newtonian potential}\label{sec:Newtonian}

It is challenging to correctly account for the GR contributions to the gradiometer phase shift induced by a moving point mass, as a fully covariant treatment is often algebraically involved, and unfeasible for complicated configurations such as the LMT sequence. However, with the simplified formalism derived in Sec.~\ref{sec:framework}, we can obtain the leading order $\mathcal{O}(hv)$ contributions to the gradiometer phase shift arising from a slow-varying and weak Newtonian potential in a tractable manner, which can be used to model Newtonian noises from passing planes or trains, as well as to study transient signals from new physics.

In the presence of a weak Newtonian potential $\Phi$ and in the rest frame of the source, the metric in harmonic gauge takes the form
\begin{equation}\label{eq:metric_static}
\begin{aligned}
    ds^2 =
     -(1+2\Phi)dt^2 + (1-2\Phi) dx_i dx^i
    \, ,
\end{aligned}
\end{equation}
where $i=1,2,3$ and we make use of the Einstein summation convention. In a frame where the source is moving at a constant velocity 
$\Vec{v}_s $ with its amplitude
$v_s \ll 1$, 
to leading order %
the metric takes the form  
\begin{equation}\label{eq:metric_dynamic}
\begin{aligned}
    ds^2 &= -(1+2\Phi)dt^2 - 8 \Phi {v_{s,i}} dx^i dt +  (1-2\Phi)
    dx_i dx^i
    \, ,
\end{aligned}
\end{equation}
which differs from Eq.~\eqref{eq:metric_static} in so far as $h_{0i} = h_{i0} =-4\Phi {v_{s,i}} \neq 0$. For a derivation, see~Appendix~\ref{sec:geodesic}. In the weak field limit, potentials that are sourced by different bodies can be added linearly. This allows us to conveniently choose an inertial frame at rest in the Earth's gravitational field. %

Unlike signals from periodic sources (e.g., GW, ultralight bosonic fields), transient signals from a moving Newtonian potential do not factorize into signal strength and response function in the time domain, making it challenging to understand the signal frequency dependence and analyze the detector sensitivity. Therefore, we directly work in the Fourier domain, where the factorization of the detector response is more straightforward.
As the source of the Newtonian potential can move in a general direction, it is necessary to express the result in three spatial dimensions. %
For simplicity, in this section, we also assume that the two AIs are located at two ends of the baseline, i.e.~$\Vec{x}_{\mathrm{AI}_1} = \Vec{x}_\mathrm{L}$ for the MZ sequence, and $\Vec{x}_{\mathrm{AI}_1} = \Vec{x}_{\mathrm{L}_1} $, $\Vec{x}_{\mathrm{AI}_2} =\Vec{x}_{\mathrm{L}_2}$ for the LMT sequence. The direction of the baseline is defined by $\Vec{n}$.

Let us first focus on the MZ configuration. Following~Eq.~\eqref{eq:grad-phase-MZ-FT}, the total phase shift in Fourier space reads as
\begin{widetext}
\begin{equation} \label{eq:phi_MZ_New}
\begin{split}
    \Delta \widetilde{\phi}^\mathrm{MZ}_\mathrm{grad}(\omega) = 
 \omega_a \, T^2 \, 
    K_{\rm MZ}   (\omega)
 \Bigg\{  
 & \underbrace{ - \omega^2\, \Delta \widetilde{\mathcal{T}}_\mathcal{S}^+ (\omega,\Vec{x}_{\mathrm{AI}_1}, \Vec{x}_{\mathrm{AI}_2})}_{\text{Shapiro}} \\ & 
 +
\underbrace{ \Big[ \Vec{n} \cdot \left( \nabla - \, 4 \, i \omega \, \Vec{v}_s \right) }_{\rm Doppler} \underbrace{- i \omega \Big]}_{\rm Einstein} \Big[ \widetilde{\Phi}(\omega,\Vec{x}_{\mathrm{AI}_1}) - e^{i\omega L}\widetilde{\Phi}(\omega,\Vec{x}_{\mathrm{AI}_2}) \Big]
\Bigg\}  \, .
\end{split}
\end{equation}
Here, the Shapiro time delay is expressed in terms of Eq.~\eqref{eq:TScurl} with the metric perturbation introduced in Eq.~\eqref{eq:metric_dynamic}, i.e.
\begin{align}\label{eq:Newt_Shapiro}  
       \Delta \mathcal{T}_\mathcal{S}^+(t,\Vec{x}_{\mathrm{AI}_1}, \Vec{x}_{\mathrm{AI}_2})
       &= -2 (1+ 2\, \vec{v}_s\cdot\vec{n}) \int_{-L/2}^{L/2}\Phi\left(t+\frac{L}{2}+ x',\Vec{x}_\mathrm{mid}+ x'\vec{n}\right) dx' \nonumber \, , \\
       \Delta \mathcal{T}_\mathcal{S}^-(t,\Vec{x}_{\mathrm{AI}_2}, \Vec{x}_{\mathrm{AI}_1})
       &= -2 (1- 2\, \vec{v}_s\cdot\vec{n}) \int_{-L/2}^{L/2}\Phi\left(t+\frac{L}{2}- x',\Vec{x}_\mathrm{mid}+ x'\vec{n}\right) dx'\, ,
\end{align}
where we have changed the integration variable and defined $\Vec{x}_\mathrm{mid} \equiv (\Vec{x}_{\mathrm{AI}_1} + \Vec{x}_{\mathrm{AI}_2})/2 $ as the midpoint of the baseline.\footnote{This quantity is equivalent to Eq.~\eqref{eq:TScurl}.}
Inspecting the frequency dependence in~Eq.~\eqref{eq:phi_MZ_New}, at low frequencies (i.e. $\omega T,~\omega L \ll 1$), the signal is dominated by the difference in the gradient of the potential evaluated at the location of the two AIs, i.e. the acceleration gradient. Recalling the asymptotic behavior of the $K_\mathrm{MZ}$,  
the total phase shift is approximately
\begin{equation}
\begin{split}
 \Delta \widetilde{\phi}^\mathrm{MZ}_\mathrm{grad}(\omega)%
 \approx \omega_a T^2  \, \Vec{n} \cdot  \left[ \nabla \widetilde{\Phi} (\omega,\mathbf{x}_{\rm AI_1}) - 
 \nabla \widetilde{\Phi} (\omega,\mathbf{x}_{\rm AI_2}) \right] ,
 \end{split}
\end{equation}
which agrees with the familiar leading order non-relativistic phase shift~\cite{PhysRevLett.67.181}.

Similarly, for an LMT sequence, following~Eq.~\eqref{eq:LMT_tot_F}, the total phase shift in the Fourier space induced by a slow-varying Newtonian potential reads as
\begin{equation} \label{eq:point_source_phase}
\begin{split}
  & \Delta \widetilde{\phi}^\mathrm{LMT}_\mathrm{grad}(\omega) =  \frac{1}{2}
  k_{\rm eff} \, T^2 \, 
    K_{\rm MZ}   (\omega)
 \Bigg\{  
 \underbrace{ - \omega^2\,
 \Big[ K_{\rm LMT}^+ \Delta \widetilde{\mathcal{T}}_\mathcal{S}^+ (\omega, \Vec{x}_{\mathrm{AI}_1}, \Vec{x}_{\mathrm{AI}_2})
 + K_{\rm LMT}^- \Delta \widetilde{\mathcal{T}}_\mathcal{S}^- (\omega, \Vec{x}_{\mathrm{AI}_2}, \Vec{x}_{\mathrm{AI}_1})
 \Big]
 }_{\text{Shapiro}} \\
& \qquad \qquad\qquad\qquad
\underbrace{ - i \omega\,
 \Big[ \big( K_{\rm LMT}^+ - e^{i\omega L}K_{\rm LMT}^- \big) \widetilde{\Phi}(\omega, \Vec{x}_{\mathrm{AI}_1})
 - \big( K_{\rm LMT}^+ - e^{- i\omega L} K_{\rm LMT}^- \big) e^{i \omega L} \widetilde{\Phi}(\omega, \Vec{x}_{\mathrm{AI}_2})
 \Big]
 }_{\text{Einstein}} \\
 & \  \
\underbrace{ 
+ \Vec{n} \cdot \left( \nabla - \, 4 \, i \omega \, \Vec{v}_s \right) \,
 \Big[ \big( K_{\rm LMT}^+ + e^{i\omega L} K_{\rm LMT}^- \big) \widetilde{\Phi}(\omega, \Vec{x}_{\mathrm{AI}_1})
 - \big( K_{\rm LMT}^+ + e^{-i\omega L} K_{\rm LMT}^- \big) e^{i \omega L} \widetilde{\Phi}(\omega, \Vec{x}_{\mathrm{AI}_2})
 \Big] 
 }_{\text{Doppler}}  \Bigg\} \,,
\end{split}
\end{equation}
where it is understood that $K_{\rm LMT}^{\pm}$ are functions of frequency as given in~Eq.~\eqref{eq:KLMTall} and $\nabla \widetilde{\Phi}(\omega, \vec{x}_\mathrm{AI_{1,2}}) \equiv \nabla \widetilde{\Phi}(\omega, \vec{x}) |_{\vec{x}_\mathrm{AI_{1,2}}}$. Notice that
\begin{equation}
\begin{split}
\lim_{\omega \to 0} K_{\rm LMT}^{+} (\omega) = 1 - \frac{(n-2)L}{T},\quad \lim_{\omega \to 0} K_{\rm LMT}^{-} (\omega) = 1 - \frac{nL}{T}.
\end{split}
\end{equation}
At low frequencies (i.e. $\omega T,~n\omega  L \ll 1$), the phase shift is therefore dominated by 
\begin{equation}\label{eq:dphi_LMT_grad_low_omega}
\begin{split}
\Delta \widetilde{\phi}^\mathrm{LMT}_\mathrm{grad}(\omega) \approx  k_{\rm eff}\,T^2\, 
 \left( 1-\frac{(n-1)L}{T} \right)
\Vec{n} \cdot  \left[ \nabla \widetilde{\Phi} (\omega,\mathbf{x}_{\rm AI1}) - 
\nabla \widetilde{\Phi} (\omega,\mathbf{x}_{\rm AI2}) \right].
 \end{split}
\end{equation}
\end{widetext}
Notice that this result differs from the MZ phase shift by a factor of $(1-(n-1)L/T)$. This can be attributed to the effective spacetime area enclosed by a single AI. In an LMT sequence, because of the incremental velocity boost, the maximum separation between atom paths is $\Delta x \propto k_\mathrm{eff} \big(T-(n-1)L\big)$ instead of $\Delta x\propto k_\mathrm{eff} T$. Hence, the enclosed spacetime area receives a correction proportional to ${(n-1)L/T}$ with respect to the MZ case. %

\begin{figure*}[t!]
	\includegraphics[width=0.95\textwidth]{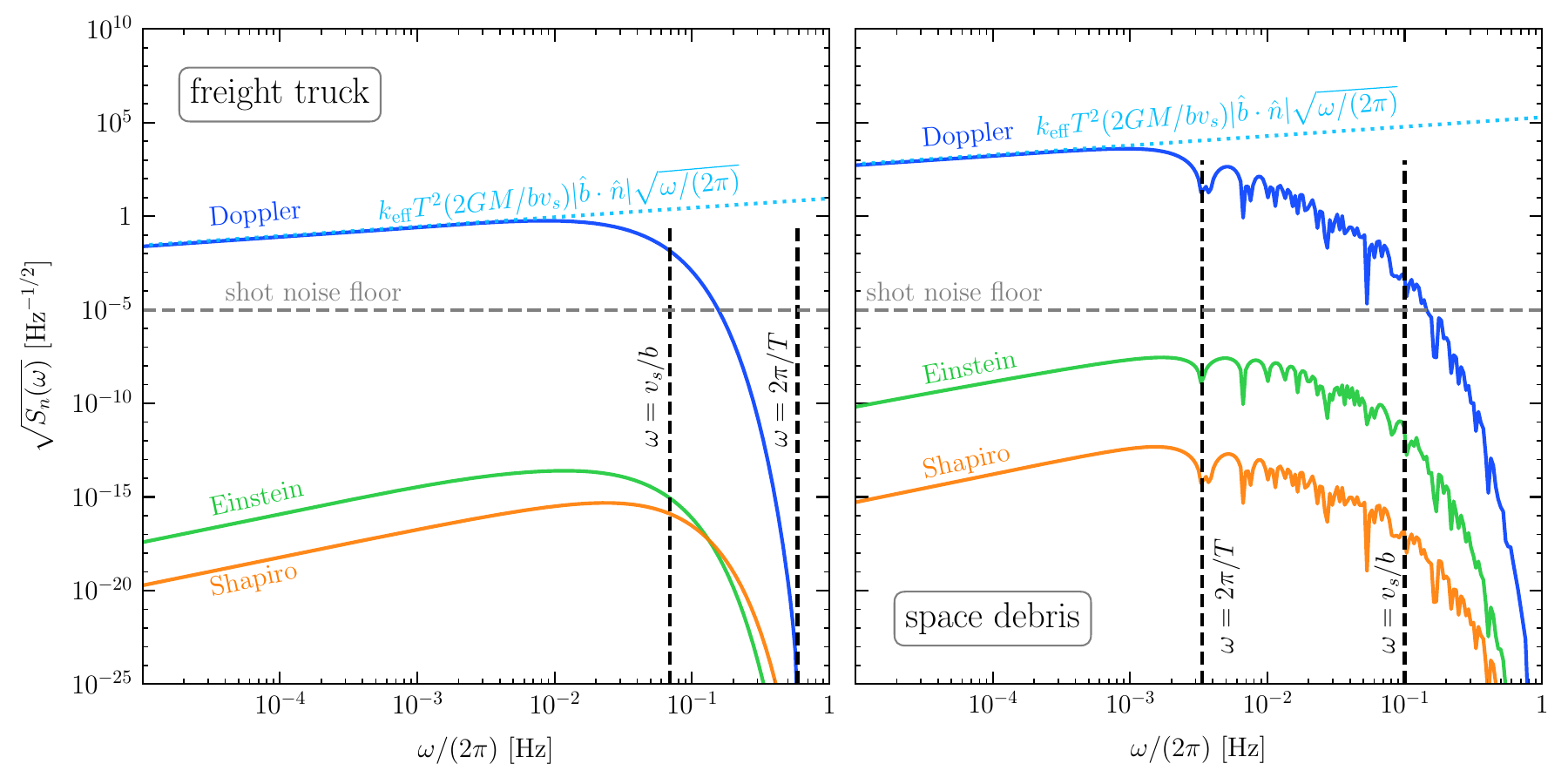} 
	\caption{The square root of the power spectral density, $\sqrt{S_n(\omega)}\equiv\sqrt{\omega/2\pi}|\Delta\phi_{\mathrm{grad}}(\omega)|$, induced by a moving Newtonian noise source. The left panel shows a freight truck with mass $M = 10^3$~kg, velocity $v_s = 25$~km/h, and impact parameter $b = 100$~m, moving near a terrestrial, single-photon atom gradiometer (AG) utilizing the large-momentum-transfer (LMT) pulse sequence with baseline $L = 1$~km, interrogation time $T = 1.7$~s, and $n = 2500$. Experimental parameters align with advanced designs potentially used in future vertical gradiometers such as AION-100, MAGIS-100, and AION-km (see Table~1 in Ref.~\cite{Badurina:2022ngn}). The right panel depicts a piece of space debris with mass $M = 10^4$~kg, velocity $v_s = 10$~m/s, and impact parameter $b = 100$~m, passing by a space-based AG, also utilizing the LMT sequence, with baseline $L = 4.4\times 10^7$~m, interrogation time $T = 150$~s, and $n = 250$. These parameters match the proposed AEDGE experiment~\cite{AEDGE:2019nxb}. Both setups employ the clock transition ($5$$s^2 $$^1$S$_0$$\leftrightarrow$5s5p$^3$P$_0$) in Sr-87 with an angular transition frequency $\omega_a = 2.70 \times 10^{15}$~Hz. The source's impact parameter spans an angle of $45^\circ$ with the baseline. Individual contributions from the Einstein (green), Doppler (blue), and Shapiro (orange) effects, as derived in Eq.~\eqref{eq:point_source_phase}, are illustrated. The estimated level of shot noise, $\sqrt{S^{(\mathrm{shot})}_n(\omega)}= 10^{-5}$~Hz$^{-1/2}$, is plotted in grey for comparison. The low-frequency behaviors of the dominant contributions are shown in sky blue. Finally, the characteristic frequencies of the noise spectrum and the response functions, $\omega = v_s/b$ and $\omega = 2\pi/T$, are plotted in black for reference.}\label{fig:spectrum} %
\end{figure*}

To better understand the impact of Newtonian noise on the spectrum of an AG experiment, we use the results from this section to compute the power spectral density of the gradiometer phase, which is
defined as $S_n(\omega)\equiv \omega/(2\pi)|\Delta \widetilde{\phi}_{\mathrm{grad}}(\omega)|^2$~\cite{Moore:2014lga}. As an example, in the left panel of~Fig.~\ref{fig:spectrum}, we plot the square root of the power spectral density (PSD) induced by a 
freight truck with a mass $M = 10^3\,$kg, an impact parameter $b = 100\,$m from the nearest AI, and a constant velocity $v_s = 25\,$km/h. Experimental parameters are derived from~Table 1 in \cite{Badurina:2022ngn}
to align with advanced designs that are potentially achievable by future experiments employing the LMT pulse sequence, such as AION-km. The estimated shot noise level, $S^{(\mathrm{shot})}_n(\omega)$, is shown for comparison. The contributions from Doppler, Shapiro and Einstein term are individually displayed. If one treats the Newtonian noise as a `signal" to be compared against the AG shot noise, then the ``signal-to-noise ratio" (SNR) can be expressed as~\cite{Moore:2014lga}
\begin{equation} \label{eq:SNR_def}
    \mathrm{SNR}^2 = \int_{\log\omega_{\min}}^{\log\omega_{\max}} d\log\omega\frac{4 S_n(\omega)}{S^{(\mathrm{shot})}_n(\omega)} \, ,
\end{equation}
where $\omega_{\min}$ and $\omega_{\max}$ are the minimum and maximum angular frequencies of the AG's sensitivity band, respectively. Hence, the SNR can be quickly estimated by eye by integrating the ratio of individual contributions to the shot-noise level over logarithmic frequencies. Other noise sources with non-trivial frequency dependence (i.e. colored noise) can also be incorporated by suitable modifications to Eq.~\eqref{eq:SNR_def} and the noise curves in Fig.~\ref{fig:spectrum}. As expected, the spectrum is dominated by the Doppler term, which far exceeds the noise floor for $\omega \lesssim 2\pi \times 10^{-2}$~Hz and would therefore impact a terrestrial experiment's projected reach, e.g., to mid-frequency GWs. %
At low frequencies, the dominant phase shift contribution of the Doppler phase shift originates from the first AI as $|\Delta\phi_{\mathrm{grad}}|\sim k_{\rm{eff}}T^2(2GM/b v_s)|\unit{b}\cdot\unit{n}|$, where $(2GM/bv_s)$ is the characteristic amplitude of the Fourier transform of the acceleration caused by a transiting source with impact parameter $b$, as indicated in Fig.~\ref{fig:spectrum}, where we show the analytic scaling with a cyan dotted curve. At $\omega\sim v_s/b$, the Shapiro contribution is the most subdominant and is parametrically suppressed by $v_s^2$ compared to the Doppler term due to its nature as a relativistic correction. Finally, at $\omega\sim v_s/b$, the Einstein contribution is parametrically suppressed by $\sim v_s(L/T)$ compared to the Doppler term. All contributions scale as power laws at low frequency, and are exponentially suppressed at high frequency above $\omega\sim v_s/b$, which is the (inverse) characteristic time-scale of the moving source.

It is worth exploring whether the Einstein and Shapiro terms can surpass the shot noise floor in a more futuristic setup. In the right panel of Fig.~\ref{fig:spectrum}, we show the PSD for a piece of space debris with mass $M = 10^4$ kg, an impact parameter $b = 100$ m from the nearest AI, and a constant velocity $v_s = 10$ m/s relative to a space-based AG. The small relative velocity could arise from both the debris and the AG orbiting at similar altitudes above Earth's surface. Experimental parameters are chosen to match the AEDGE proposal~\cite{AEDGE:2019nxb}. As indicated by Eq.~\eqref{eq:point_source_phase}, the Einstein spectrum peaks at $\omega \sim v_s/b$, with $\Delta\phi_{\mathcal{E}}\sim 2k_{\mathrm{eff}}LT\omega(GM/v_s)$. An estimate with this formula naively suggests that the Einstein contribution might exceed the noise floor for certain realistic Newtonian noise parameters. However, the 
response kernels $K_{\mathrm{MZ}}(\omega)$ and $K^{\pm}_{\mathrm{LMT}}(\omega)$ as defined in Eq.~\eqref{eq:MZ-response} and Eq.~\eqref{eq:KLMTall} introduce additional suppression factors of $(\omega T)^{-1}$ for frequencies beyond $\omega\sim 2\pi/T$, which suppresses all contributions at the naively expected peak spectrum frequency. The $\sin(\omega T)$ term in the response functions is also responsible for the choppiness in the shape of the spectrum for frequencies $\omega \gtrsim 2\pi/T$. Consequently, it is unlikely that the Einstein contribution to a realistic Newtonian noise will be significant in any experimental search. A similar conclusion can be drawn for the Shapiro contribution, which, although not suppressed by factors of $L/T$, is further suppressed by a factor of $v_s$. It is important to note that while noise sources with a larger $v_s$ could be considered, the peak frequency of these sources is greater than $2\pi/T$, thereby further suppressing the noise amplitude, and thus can be neglected.

\section{Discussion \& Conclusions}
\label{sec:disc_conc}

In this work, we have developed a simplified gauge-invariant formalism for calculating the leading-order gravitational phase shift at $\mathcal{O}(hv)$ in a single-photon atom gradiometer under the influence of a generic metric perturbation. We have found that the leading-order contributions can be fully accounted for by the rest mass correction of the atom excited states and can be conveniently written in terms of Doppler, Shapiro, and Einstein time delays in analogy with laser interferometers. 

We applied this formalism to MZ and LMT gradiometer configurations, and we calculated the gravitational signals from GWs and slow-varying weak Newtonian potentials in these experiments. Our calculations of the GW phase shift are performed in both the TT gauge and the proper detector frame. Our result confirms the gauge-invariance of the leading-order phase shift and agrees with the literature \cite{Graham:2012sy,Graham:2016plp}. In the proper detector frame, the Doppler phase shift gives the familiar $\Delta\phi\propto k_\mathrm{eff}aT^2$ with an acceleration induced by GW in the long wavelength limit $\omega_\mathrm{g} L\ll 1$. We also derived the signal of a slow-varying weak Newtonian potential in the frequency domain and calculated the response functions for both the MZ and LMT configurations. We found that in the laser's rest frame, the dominant effect comes from the Doppler phase shift, which gives the $k_\mathrm{eff}a T^2$ scaling in the low frequency regime; the Einstein contribution receives a $v_s L/T$ suppression. 

The formalism presented in this paper is valid to $\mathcal{O}(hv)$. Extending the range of validity to higher order in the atom velocity and the metric perturbation requires a punctilious analysis, which we leave to future work. For instance, to calculate effects at $\mathcal{O}(v^2)$ or $\mathcal{O}(h^2)$, one would need a more careful and precise treatment of the separation phase and atom paths, as well as the momentum transferred in atom-light interactions. In our derivation, we assumed that the path $C$ is closed (i.e. the worldlines of the atomic wavepacket's c.o.m. intersect at the null hypersurface defining the measurement); to achieve this in the laboratory, prior knowledge of the metric is required. This, however, is not possible for the type of searches envisaged here. Consequently, in general the atom's worldlines are not expected to overlap at the measurement port. Let us refer to this path as $C_\mathrm{true}$. In this case, the phase shift for each AI in the gradiometer set-up receives a correction $\Delta \phi_\mathrm{separation} = -\bar{p}_\mu \Delta \mathcal{X}^\mu$ commonly known as the separation phase~\cite{Dimopoulos:2008hx}. Here, $\bar{p}_\mu$ is the average four momentum of the two wavepackets being recombined at a particular measurement port and $\Delta \mathcal{X}^\mu$ is the coordinate separation between the right and left wavepackets on the null hypersurface defining the measurement. Crucially, $\Delta \mathcal{X}^\mu$ enters at order $\mathcal{O}(hv)$, while $\bar{p}_\mu$ is non-vanishing at zeroth order in both the metric perturbation and the atom velocity. This observation would suggest that this correction should be considered in our framework. Nevertheless, as we show in Appendix~\ref{app:sep_phase}, the difference between the propagation phase shift for $C$ and the sum of the propagation and separation phase shifts for the true path $C_\mathrm{true}$ is zero up to corrections that enter at higher order. 
Furthermore, the corrections to the atom's initial conditions in $\widetilde{C}$ also enter the observable at $\mathcal{O}(h v^2)$, as do the corrections to the atom recoil velocities due to the photons' propagation in curved spacetime and the initial launch velocity of the atoms. Finally, the equivalence between the phase shift evaluated over $C$ and $\widetilde{C}$ only holds at leading order in the metric, so an extension of our formalism to $\mathcal{O}(h^2)$ would require a careful treatment order by order in perturbation theory. 

Although we have restricted our attention to vertical gradiometers employing single-photon transitions, our results can also be modified to predict the observable in AG experiments which utilize horizontal baselines and two-photon transitions (e.g., MIGA, ELGAR and ZAIGA). In these types of experiments, the baseline is perpendicular to the initial launch velocity of the atoms and the transition from ground to excited state (and vice versa) is driven by two counter-propagating laser beams which interact with a particular atomic wavepacket almost synchronously. As a result, the emission times of two pulses are interrelated. However, following the same arguments presented in Appendix~\ref{app:DopplerShapiro}, the observable is not sensitive to $\mathcal{O}(h)$ corrections to the laser pulse emission times; therefore, only the unperturbed emission times would have to be carefully chosen. %

In conclusion, the formalism developed in this paper opens exciting new avenues for future studies. For example, our simplified framework, and especially the gradiometer phase shift expressions presented in this work, can assist with the development of Wiener filters in atom gradiometer experiments to mitigate Newtonian noise from nearby traffic (air crafts, trains, etc.) and well-monitored seismic activity.  Additionally, because of the generality of our framework insofar as the form of the metric perturbation is concerned, the key results of our work can be used to study signatures of new physics, %
such as macroscopic dark matter, which we explore in a companion paper~\cite{Badurina:2025xwl}, and more exotic signatures, such as violations of general coordinate invariance. 

\begin{acknowledgments}
We thank Albert Roura and Allic Sivaramakrishnan for useful discussions, and the anonymous referee for their comments.
L.B., Y.D., V.L., Y.W. and K.Z. are supported by the U.S. Department of Energy, Office of Science, Office of High Energy Physics under Award No. DE-SC0011632, and by the Walter Burke Institute for Theoretical Physics. KZ is also supported by a Simons Investigator award, and by Heising-Simons Foundation “Observational Signatures of Quantum Gravity” collaboration Grant No. 2021-2817.
\end{acknowledgments}

\appendix

\section{Separation phase and path deformations}\label{app:sep_phase}

In this appendix, we explicitly demonstrate that the propagation phase shift along the closed path
$C$, introduced in section~\ref{sec:framework_derivation}, is sufficient to compute the observable for realistic paths, i.e. when the atomic wavepackets do not perfectly recombine at the measurement port.

\begin{figure}
    \centering
    \includegraphics[width=0.35\linewidth]{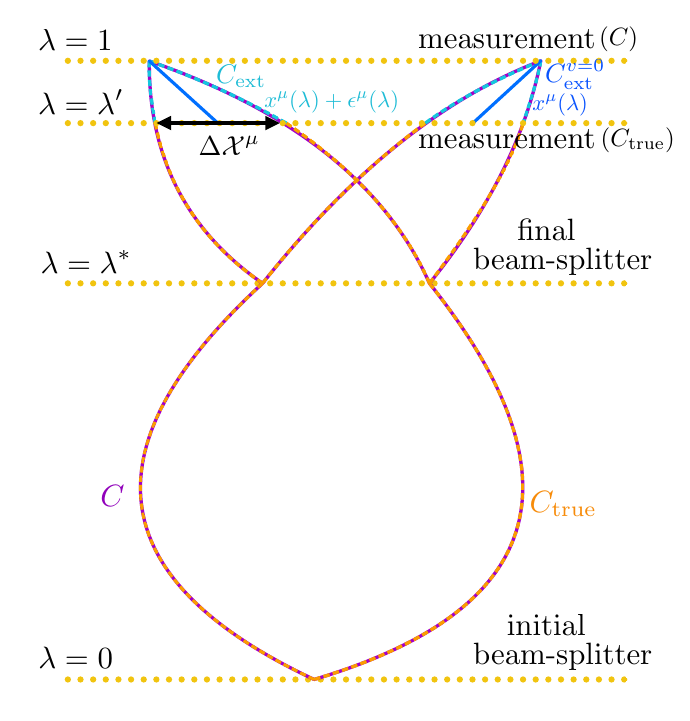}
    \caption{Schematic spacetime diagram showing the atom trajectories defined to address the separation phase. The dotted yellow lines denote hypersurfaces defined by the initial beam-splitter pulse, the final beam-splitter pulse, and the measurements for different paths. We define the atom trajectory that ends with the $\lambda=\lambda'$ measurement, and whose coordinate separation at the measurement pulse is $\Delta \mathcal{X}^\mu$, as $C_\mathrm{true}$ (dashed orange). The trajectory $C$ (solid purple) ends with the measurement pulse at $\lambda=1$. The difference between $C_\mathrm{true}$ and $C$ is defined as $C_\mathrm{ext}$ (dashed cyan), which can be deformed into $C_\mathrm{ext}^{v=0}$ (solid blue) by setting $v=0$.}
    \label{fig:app:paths}
\end{figure}

For definiteness, let $C_\mathrm{true}$ be the path taken by the atomic wavepackets when the experimentalist has no prior knowledge of the metric tensor, i.e. the \textit{true} path. In $C_\mathrm{true}$, the wave packets on the right ($\mathcal{R}$) and left ($\mathcal{L}$) arms of an AI do not intersect at either the final beamsplitter pulse or the null hypersurface defining the measurement. This can be achieved, for example, by performing the projective measurement on the atom clouds at some earlier time than that chosen in $C$. In this case, $C_\mathrm{true}$ can be related to $C$, as defined in section~\ref{sec:framework_derivation}, via $C = C_\mathrm{true} + C_\mathrm{ext}$, where $C_\mathrm{ext}$ accounts for the atom's evolution between the measurement in $C_\mathrm{true}$ and the measurement in $C$, as we show in Fig.~\ref{fig:app:paths}. Hence, the difference between the propagation phase shifts for $C$ and $C_\mathrm{true}$ can be written as 
\begin{equation}\label{app:eq:diff}
\Delta \phi - \Delta \phi_\mathrm{true} = \int_{C_\mathrm{ext}} m d\tau \Big |_\mathcal{R} -\int_{C_\mathrm{ext}} m d\tau \Big |_\mathcal{L} \, . 
\end{equation}

Let us now introduce the ancillary path $C_\mathrm{ext}^{v=0}$, which accounts for the evolution of the wavepackets in the limit of zero recoils. Let us parametrize the geodesics in $C_\mathrm{ext}^{v=0}$ and $C_\mathrm{ext}$ by $x^\mu(\lambda)$ and $x^\mu(\lambda) + \epsilon^\mu(\lambda)$, respectively, with $\epsilon^\mu(1) = 0$ and $\epsilon^\mu(\lambda')|_{\mathcal{L,R}} = \mp \Delta \mathcal{X}^\mu/2$. Note that $\Delta \mathcal{X}^\mu$ is the $\mathcal{O}(hv)$ coordinate separation between the right and left paths along the measurement null hypersurface in $C_\mathrm{true}$, and the minus sign applies to the left arm only. Since the paths in $C_\mathrm{ext}^{v=0}$ and $C_\mathrm{ext}$ are geodesics, the propagation phase accumulated along the left and right paths in $C_\mathrm{ext}$ may be rewritten as
\begin{equation}\label{eq:app:action}
\begin{aligned}
 \int_{C_\mathrm{ext}} m d\tau \Big |_\mathcal{L,R} = & \int_{C_\mathrm{ext}^{v=0}} m d\tau \Big |_\mathcal{L,R} + \left [ \frac{\partial L}{\partial \dot{x}^\mu}  \epsilon^\mu \Big |_\mathcal{L,R} \right]_{\lambda'}^{1} 
 + \mathcal{O}(\epsilon^2) \, ,
 \end{aligned}
\end{equation}
where we introduced the notation {$L = m\sqrt{-g_{\mu\nu}(x)\dot{x}^\mu \dot{x}^\nu}$} and $\dot{x}^\mu = dx^\mu/d\lambda$. 

We now make a crucial observation: since the atomic wavepackets at a particular measurement port are in the same state, the propagation phase shift accumulated in $C_\mathrm{ext}^{v=0}$ is zero. Hence, replacing $d/d\lambda$ with $L/m \times d/d\tau$, and using $\partial L/\partial \dot{x}^\mu= m^2 g_{\mu \nu} \dot{x}^\nu/L$ and Eq.~\eqref{eq:app:action}, Eq.~\eqref{app:eq:diff} may be written as
\begin{equation}\label{app:eq:diff_final}
\begin{aligned}
\Delta \phi - \Delta \phi_\mathrm{true} = -\bar{p}_\mu \Delta \mathcal{X}^\mu + \mathcal{O}(\epsilon^2) \, ,
\end{aligned}
\end{equation}
where $\overline{p}_\mu = (p_\mu |_\mathcal{R} + p_\mu |_\mathcal{L})/2$ is the average four momentum momentum at the application of the measurement pulse in $C_\mathrm{true}$.
Since $\Delta \phi_\mathrm{separation} = -\bar{p}_\mu \Delta \mathcal{X}^\mu$ is the separation phase in $C_\mathrm{true}$~\cite{Dimopoulos:2008hx}, we conclude that the propagation phase for $C$ is equivalent to the sum of the propagation and separation phase shifts for $C_\mathrm{ext}$ to $\mathcal{O}(hv)$.

\section{Shapiro and Doppler phase shifts in single-photon atom gradiometry}\label{app:DopplerShapiro}

In atom gradiometers utilizing single-photon transitions, the Shapiro and Doppler phase shifts are manifestly independent of the dynamics of the laser. This is due to the fact that a laser pulse that starts an excited state segment of one atom interferometer will also start the excited state of the other atom interferometer, and those excited state segments are ended by another common laser pulse. To see this explicitly, we consider the Doppler and Shapiro phase shifts for an arbitrary excited state path segment. %

Throughout this appendix, we use overbars to identify unperturbed quantities, and denote $\mathcal{O}(h)$ corrections to coordinates by $\delta t$ and $\delta x$. Let us first consider a gradiometer configuration with a single laser (e.g., a Mach-Zehnder gradiometer). For simplicity, let us set the baseline along the $x$-direction (i.e. $i = 1$). Let $(t^I_{\mathrm{L}},x^I_{\mathrm{L}}) = (\bar{t}^I_{\mathrm{L}}+ \delta t^I_{\mathrm{L}},\bar{x}^I_{\mathrm{L}} + \delta x^I_{\mathrm{L}})$ be the spacetime point at which the laser pulse that initialises the excited state paths in both AIs is emitted. This pulse will drive the transition from ground to excited state in the first and second AI at $(t^I_1,x^I_1) = (\bar{t}^I_{1}+\delta t^I_1,\bar{x}^I_{1} + \delta x^I_1)$ and $(t^I_2,x^I_2) = (\bar{t}^I_{2}+\delta t^I_2,\bar{x}^I_{2}+\delta x^I_2)$, respectively. Let $(t^F_{\mathrm{L}},x^F_{\mathrm{L}})=(\bar{t}^F_{\mathrm{L}}+\delta t^F_{\mathrm{L}},\bar{x}^F_{\mathrm{L}}+\delta x^F_{\mathrm{L}})$ be the spacetime point at which the laser pulse that drives the transition from excited to ground state is emitted. This second pulse will drive the transition from excited to ground state in the first and second AI at $(t^F_1,x^F_1) = (\bar{t}^F_{1}+\delta t^F_1,\bar{x}^F_{1} + \delta x^F_1)$ and $(t^F_2,x^F_2) = (\bar{t}^F_{2}+\delta t^F_2,\bar{x}^F_{2}+\delta x^F_2)$, respectively. We show this diagrammatically in the left panel of Fig.~\ref{fig:sequences}. Since photons travel on null geodesics, 
the $\mathcal{O}(h)$ corrections to the coordinate time at which the atom and laser worldlines intersect are given by 
\begin{widetext}
\begin{align}\label{eq:app:t-corr}
\delta t^{I,F}_{1,2} &= \delta t_\mathrm{L}^{I,F} + \int_{\delta x^{I,F}_{\mathrm{L}}}^{\delta x^{I,F}_{1,2}} dx'+\int_{\bar{x}^{I,F}_{\mathrm{L}} }^{\bar{x}^{I,F}_{1,2}}\mathcal{H}_{+} (\bar{t}_{\mathrm{L}}^{I,F}-\bar{x}_{\mathrm{L}}^{I,F}+x',x')dx'+\mathcal{O}(h^2) \, ,
\end{align}
where we introduced $\mathcal{H}_{+}$ in Eq.~\eqref{eq:mathcal_H} with $\vec{n}=\unit{x}$. Note the appearance of the $\mathcal{O}(h)$ correction to the spacetime points at which the photons are emitted, which in turn gives rise to the $\Delta t_\mathrm{laser}^{(k)}$ term in Eq.~\eqref{eq:Shapiro_schematic}. From Eq.~\eqref{eq:grad-form}, the sum of the Shapiro and Doppler phase shifts for this particular excited state path segment is given by
\begin{equation}\label{eq:app:dopp-shap-k}    \Delta\phi^{(\star)}_{\mathrm{grad},\mathcal{D}}+\Delta\phi^{(\star)}_{\mathrm{grad},\mathcal{S}} = \omega_a\left(\delta t^F_{1}-\delta t^I_1\right)-\omega_a\left(\delta t^F_{2}-\delta t^I_{2}\right)=\omega_a \left(\delta t^I_{2}-\delta t^I_{1}\right)-\omega_a\left(\delta t^F_{2}-\delta t^F_{1}\right) \, .
\end{equation}
Combining Eq.~\eqref{eq:app:t-corr} with Eq.~\eqref{eq:app:dopp-shap-k}, it immediately follows that the $\delta t_L$ and $\delta x_L$ terms cancel out (i.e. the gradiometer observable is insensitive to $\Delta t_\mathrm{laser}^{(*)}$). Furthermore, from the linearity of integration, it also follows that the integration limits do not depend on $\delta x_L$. Explicitly, %
\begin{equation}\label{eq:app:step1}
\begin{aligned}
\delta t_2^{I,F}-\delta t_1^{I,F} %
& = \int_{\delta x^{I,F}_1}^{\delta x^{I,F}_{2}} dx'+\int_{\bar{x}^{I,F}_{1} }^{\bar{x}^{I,F}_{2}}\mathcal{H}_{+} (\bar{t}_{\mathrm{L}}^{I,F}-\bar{x}_{\mathrm{L}}^{I,F}+x',x')dx' \, .
\end{aligned}
\end{equation}
After some algebra, the terms of the LHS of Eq.~\eqref{eq:app:dopp-shap-k} may be rewritten using Eq.~\eqref{eq:app:step1} as 
\begin{equation} \label{eq:app:final-MZ}
\begin{aligned}
\Delta\phi^{(\star)}_{\mathrm{grad},\mathcal{D}} &= %
\omega_a \left (\int_{\delta x^{I}_1}^{\delta x^{I}_{2}} dx' - \int_{\delta x^{F}_1}^{\delta x^{F}_{2}} dx' \right )
= \omega_a \left ( \delta x^{I}_2  + \delta x^{F}_1 - \delta x^{I}_1 - \delta x^{F}_2 \right ) \, , \\ %
\Delta\phi^{(\star)}_{\mathrm{grad},\mathcal{S}} &= 
\omega_a \left (\int_{\bar{x}^{I}_{1} }^{\bar{x}^{I}_{2}}\mathcal{H}_{+} ( \bar{t}_{\mathrm{L}}^{I}-\bar{x}_{\mathrm{L}}^{I}+x',x')dx' - \int_{\bar{x}^{F}_{1} }^{\bar{x}^{F}_{2}}\mathcal{H}_{+} (\bar{t}_{\mathrm{L}}^{F}-\bar{x}_{\mathrm{L}}^{F}+x',x')dx' \right ) \, .
\end{aligned}
\end{equation}
In this example, we assumed that the excited state path segment is on the right arm of the interferometer. If this segment were on the left arm of the interferometer, one can easily show that the Doppler and Shapiro would gain an overall minus sign.

\begin{figure*}
	\includegraphics[width=0.37\textwidth]{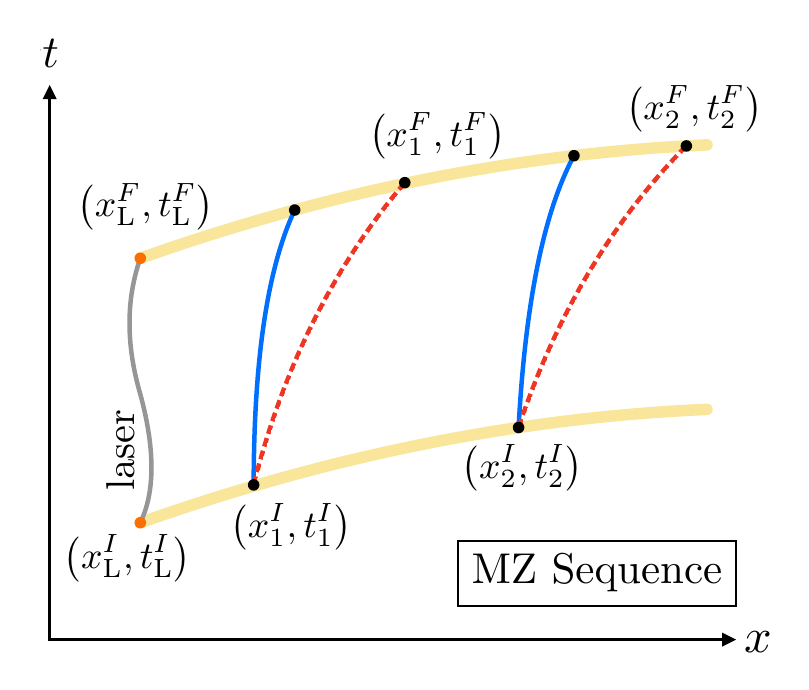} 
     \includegraphics[width=0.37\textwidth]{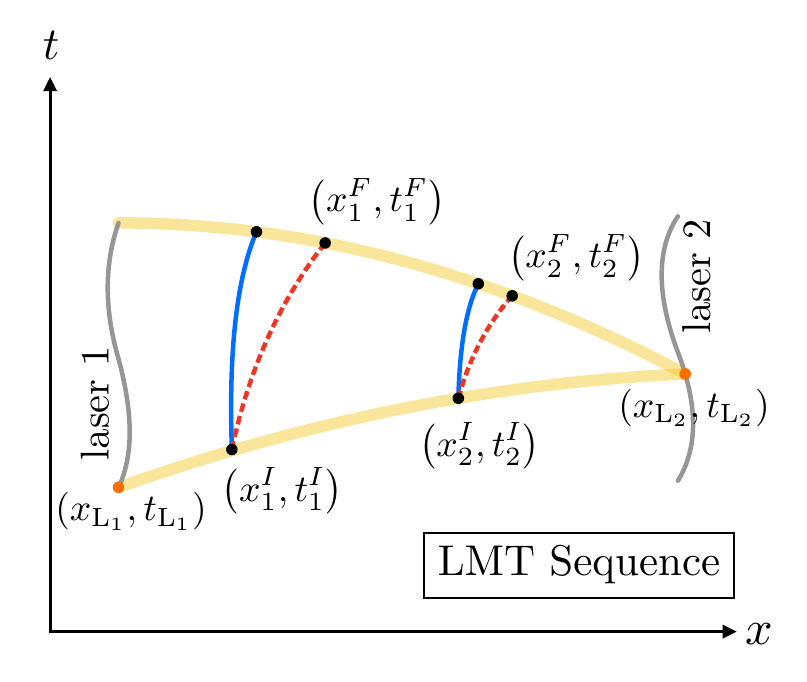}
	\caption{Schematic representation of an arbitrary excited state path segment (dashed red line) in a single-photon atom gradiometer. On the left, we show the relevant spacetime diagram for a MZ gradiometer sequence, where the transitions are driven by a single laser source. On the right, we show the relevant spacetime diagram for a LMT gradiometer sequence, where the transitions are driven by two laser sources}\label{fig:sequences}
\end{figure*}

Let us now show the same result for a configuration employing two laser sources (e.g., a LMT gradiometer configuration). Let $(t_{\mathrm{L}_1},x_{\mathrm{L}_1}) = (\bar{t}_{\mathrm{L}_1}+ \delta t_{\mathrm{L}_1},\bar{x}_{\mathrm{L}_1} + \delta x_{\mathrm{L}_1})$ be the spacetime point at which the first laser pulse that initialises the excited state paths in both AIs is emitted. This pulse will drive the transition from ground to excited state in the first and second AIs at spacetime points $(t^I_1,x^I_1) = (\bar{t}^I_{1}+\delta t^I_1,\bar{x}^I_{1} + \delta x^I_1)$ and $(t^I_2,x^I_2) = (\bar{t}^I_{2}+\delta t^I_2,\bar{x}^I_{2}+\delta x^I_2)$, respectively. Similarly, let $(t_{\mathrm{L}_2},x_{\mathrm{L}_2})=(\bar{t}_{\mathrm{L}_2}+\delta t_{\mathrm{L}_2},\bar{x}_{\mathrm{L}_2}+\delta x_{\mathrm{L}_2})$ be the spacetime point at which the second laser pulse that drives the transition from excited to ground state is emitted, so that the spacetime endpoints of the excited state paths of the AIs are $(t^F_1,x^F_1) = (\bar{t}^F_{1}+\delta t^F_1,\bar{x}^F_{1} + \delta x^F_1)$ and $(t^F_2,x^F_2) = (\bar{t}^F_{2}+\delta t^F_2,\bar{x}^F_{2}+\delta x^F_2)$. We show this diagrammatically in the right panel of Fig.~\ref{fig:sequences}. 
The $\mathcal{O}(h)$ corrections to the coordinate time at which the atom and laser worldlines intersect now depend on whether the pulse was emitted from the first or second laser pulse. Explicitly,
\begin{equation}
\begin{aligned}\label{eq:app:t-corr-lmt}
\delta t^{I}_{1,2} &= \delta t_{\mathrm{L}_1} + \int_{\delta x_{\mathrm{L}_1}}^{\delta x^{I}_{1,2}} dx'+\int_{\bar{x}_{\mathrm{L}_1} }^{\bar{x}^{I}_{1,2}}\mathcal{H}_{+} (\bar{t}_{\mathrm{L}_1}-\bar{x}_{L_{1}}+x',x')dx'+\mathcal{O}(h^2) \, , \\
\delta t^{F}_{1,2} &= \delta t_{\mathrm{L}_2} - \int_{\delta x_{\mathrm{L}_2}}^{\delta x^{F}_{1,2}} dx'-\int_{\bar{x}_{\mathrm{L}_2} }^{\bar{x}^{F}_{1,2}}\mathcal{H}_{-} (\bar{t}_{\mathrm{L}_2}+\bar{x}_{\mathrm{L}_2}-x',x')dx'+\mathcal{O}(h^2) \, .
\end{aligned}
\end{equation}
Making use of Eq.~\eqref{eq:app:t-corr-lmt} to write 
\begin{equation}
\begin{aligned}
\delta t_2^{I}-\delta t_1^{I} 
& = \int_{\delta x^{I}_1}^{\delta x^{I}_{2}} dx'+\int_{\bar{x}^{I}_{1} }^{\bar{x}^{I}_{2}}\mathcal{H}_{+} (\bar{t}_{\mathrm{L}_1}-\bar{x}_{\mathrm{L}_1}+x',x')dx'  \, , \\
\delta t_2^{F}-\delta t_1^{F} 
& = -\int_{\delta x^{F}_1}^{\delta x^{F}_{2}} dx'-\int_{\bar{x}^{F}_{1} }^{\bar{x}^{F}_{2}}\mathcal{H}_{-} (\bar{t}_{\mathrm{L}_2}+\bar{x}_{\mathrm{L}_2}-x',x')dx' \, ,
\end{aligned}
\end{equation}
the Doppler and Shapiro gradiometer phase shift terms (\textit{cf}. Eq.~\eqref{eq:app:dopp-shap-k}) for this path segment can be rewritten as %
\begin{equation}\label{eq:app:grad_LMT}
\begin{aligned}
\Delta\phi^{(\star)}_{\mathrm{grad},\mathcal{D}} &= %
\omega_a \left( \delta x_2^I + \delta x_2^F - \delta x_1^I - \delta x_1^F \right ) \, , \\
\Delta\phi^{(\star)}_{\mathrm{grad},\mathcal{S}} &= %
\omega_a \left (\int_{\bar{x}^{I}_{1} }^{\bar{x}^{I}_{2}}\mathcal{H}_{+} (\bar{t}_{\mathrm{L}_1}-\bar{x}_{\mathrm{L}_1}+x',x')dx' +  \int_{\bar{x}^{F}_{1} }^{\bar{x}^{F}_{2}}\mathcal{H}_{-} (\bar{t}_{\mathrm{L}_2}+\bar{x}_{\mathrm{L}_2}-x',x')dx' \right ) \, .
\end{aligned}
\end{equation}

Here, we assumed that the excited state path segment is on the right arm of the AI. The Doppler and Shapiro contributions from left-arm paths gain an overall minus sign. Furthermore, we assumed that the first pulse originated from the end of the baseline closest to the first AI. When the first pulse originates from the opposite end of the baseline, %
the Doppler and Shapiro phase shifts are given by
\begin{equation}\label{eq:app:grad_LMT_2}
\begin{aligned}
\Delta\phi^{(\star)}_{\mathrm{grad},\mathcal{D}} &= %
- \omega_a \left (\delta x_2^I + \delta x_2^F - \delta x_1^I - \delta x_1^F  \right ) \, , \\
\Delta \phi_{\mathrm{grad},\mathcal{S}}^{(\star)}  &=  - \omega_a \left (\int_{\bar{x}^{I}_{1} }^{\bar{x}^{I}_{2}}\mathcal{H}_{-} (\bar{t}_{\mathrm{L}_2}+\bar{x}_{\mathrm{L}_2}-x',x')dx' + \int_{\bar{x}^{F}_{1} }^{\bar{x}^{F}_{2}}\mathcal{H}_{+} (\bar{t}_{\mathrm{L}_1}-\bar{x}_{\mathrm{L}_1}+x',x')dx'  \right ) \, ,
\end{aligned}
\end{equation} 
for right-arm paths. The contribution from left-arm segments differs by a relative sign.
Note, also, that Eqs.~\eqref{eq:app:final-MZ}, \eqref{eq:app:grad_LMT} and \eqref{eq:app:grad_LMT_2} can be expressed in terms of the gradiometer time delays introduced in sections~\ref{sec:framework}. Indeed, the $\omega_a$-independent piece in Eq.~\eqref{eq:app:final-MZ} is the Doppler/Shapiro MZ gradiometer time delay, while the $\omega_a$-independent piece in Eqs.~\eqref{eq:app:grad_LMT} and \eqref{eq:app:grad_LMT_2} is the Doppler and Shapiro LMT gradiometer time delay of ``$>$" or ``$<$" excited state path segments, respectively (\textit{cf}.~section~\ref{sec:framework_derivation:definite_pulses}).

\section{Derivation of the phase shift formulae in the frequency domain}\label{sec:deriv}

In this appendix, we derive the form of the gradiometer phase shift inthe frequency domain. We specialise to the MZ and LMT configurations defined in section~\ref{sec:framework_derivation:definite_pulses}.

\subsection{Mach-Zehnder gradiometer}\label{app:MZ-config}

As we showed in section~\ref{sec:MZ}, the Einstein, Doppler and Shapiro gradiometer phase shift contributions for a MZ gradiometer initiated at time $t_0$ may be expressed as 
\begin{equation}\label{eq:app_grad-phase-MZ}
\begin{aligned}
\Delta \phi^\mathrm{MZ}_{\mathrm{grad},\mathcal{E,D,S}} (t_0) & = 
 \omega_a \Big( \Delta t_{\mathrm{grad},\mathcal{E,D,S}}^\mathrm{MZ} (t_0) - \Delta t_{\mathrm{grad},\mathcal{E,D,S}}^\mathrm{MZ} (t_0 + T ) \Big) \, ,
\end{aligned}
\end{equation}
where the time delays are defined as
\begin{equation} \label{eq:app:grad-time delays-MZ}
\begin{aligned}
\Delta t_{\mathrm{grad},\mathcal{E}}^\mathrm{MZ} (t) & = -\frac{1}{2}\left [\int_{t}^{t+T}  h_{00} (t',x_{\mathrm{AI}_1}) dt' - \int_{t +L }^{t + T + L}h_{00}(t',x_{\mathrm{AI}_2}) dt' \right] \, , \\ 
\Delta t_{\mathrm{grad},\mathcal{D}}^\mathrm{MZ}(t) & = n_i \Big [ \delta x^i(t+T,x_{\mathrm{AI}_1})- \delta x^i(t,x_{\mathrm{AI}_1}) +\delta x^i(t+L,x_{\mathrm{AI}_2})  -  \delta x^i(t+T+L,x_{\mathrm{AI}_2}) \Big ] \, , \\
\Delta t_{\mathrm{grad},\mathcal{S}}^\mathrm{MZ}(t) &= 
\int_{x_{\mathrm{AI}_1}}^{x_{\mathrm{AI}_2}}  \mathcal{H}_{+} (t(x'),x')dx' - \int_{x_{\mathrm{AI}_1}}^{x_{\mathrm{AI}_2}} \mathcal{H}_{+} (t(x')+T,x')dx' \, , 
\end{aligned}
\end{equation}
with $t(x')\equiv t+x'-x_{\mathrm{AI}_1}$ in the limit where the distance between the laser source and the first AI can be neglected.
Using the Fourier transform convention 
\begin{equation}
\widetilde{x}(\omega) = \int_{-\infty}^{\infty} \, dt \, x(t) \, e^{-\mi \omega t} \, ,
\end{equation}
and performing the time shift $t_0 \rightarrow t_0-T$, %
each gradiometer phase shift contribution may be expressed in the frequency domain as
\begin{equation}
\Delta \widetilde{\phi}^\mathrm{MZ}_{\mathrm{grad},\mathcal{E,D,S}} (\omega) = 
 \omega_a \Big(1 - e^{\mi \omega T} \Big) \Delta \widetilde{t}_{\mathrm{grad},\mathcal{E,D,S}}^\mathrm{MZ} (\omega) \, .
\end{equation}
Similarly, the Fourier transform of the Doppler gradiometer time delay can be expressed as
\begin{equation}
\begin{aligned}\label{eq:app:td-Doppler}
\Delta \widetilde{t}_{\mathrm{grad},\mathcal{D}}^\mathrm{MZ} (\omega) %
& = \left (e^{ \mi \omega T} - 1\right ) \left [ n_i \delta \widetilde{x}^i (\omega,x_\mathrm{AI}) e^{ \mi \omega (x_\mathrm{AI}-x_\mathrm{AI_1})} \right ]\Bigg |^{x_\mathrm{AI_1}}_{x_\mathrm{AI_2}} \, .
\end{aligned}
\end{equation}
For convenience, let us rewrite the Shapiro gradiometer time delay in terms of the Shapiro time delay computed between the laser located at $x_L$ and a particular AI (\textit{cf}. Eq.~\eqref{eq:TScurl}). Explicitly,
\begin{equation}
\Delta t^\mathrm{MZ}_{\mathrm{grad},\mathcal{S}}(t_0) = \left [\Delta \mathcal{T}^+_\mathcal{S}(t_0+T,x_\mathrm{AI_1},x_\mathrm{AI}) \right ]\Bigg |^{x_\mathrm{AI_1}}_{x_\mathrm{AI_2}} -\left [\Delta \mathcal{T}^+_\mathcal{S}(t_0,x_\mathrm{AI_1},x_\mathrm{AI}) \right ]\Bigg |^{x_\mathrm{AI_1}}_{x_\mathrm{AI_2}} \, ,
\end{equation}
with
\begin{equation}
    \Delta \mathcal{T}_\mathcal{S}^+ (t_0, x_\mathrm{AI_1}, x_\mathrm{AI})\equiv \int_{x_\mathrm{AI_1}}^{x_\mathrm{AI}}  \mathcal{H}_{+}(t_0 +(x'-x_\mathrm{AI_1}), x')dx' \, .
\end{equation}
Employing this decomposition and adopting the procedure for deriving the MZ gradiometer phase shift, the Fourier transform of the Shapiro gradiometer time delay takes the following compact form:
\begin{equation}\label{eq:app:td-Shapiro}
\Delta \widetilde{t}_{\mathrm{grad},\mathcal{S}}^\mathrm{MZ} (\omega) = \left (e^{\mi \omega T} - 1 \right ) \left [\Delta \widetilde{\mathcal{T}}^+_\mathcal{S}(\omega,x_\mathrm{AI_1},x_{\mathrm{AI}}) \right]^{x_\mathrm{AI_1}}_{x_\mathrm{AI_2}} \, .
\end{equation}
To express the Einstein gradiometer time delay in the frequency domain, we use the following identity:
\begin{equation}
\mathcal{FT}_{t\rightarrow \omega }\left \{ \int_{-\infty }^{t} f(t') \, dt'\right \} = \frac{\widetilde{f}(\omega)}{\mi \, \omega} + 2\pi \widetilde{f}(0) \delta (\omega) \, ,
\end{equation}
where $\delta (\omega)$ is the Dirac delta function~\cite{bracewell1978fourier}.
Consequently, for constant $t_A$ and $t_B$, we find
\begin{equation}\label{eq:app:identity}
\mathcal{FT}_{t\rightarrow \omega }\left \{ \int_{t+t_A }^{t+t_B} f(t') \, dt'\right \} = \left (\frac{\widetilde{f}(\omega)}{\mi \, \omega} + 2\pi \widetilde{f}(0) \delta (\omega) \right ) \left ( e^{\mi \omega t_B} - e^{\mi \omega t_A} \right ) =  \frac{\widetilde{f}(\omega)}{\mi \, \omega} \left ( e^{\mi \omega t_B} - e^{\mi \omega t_A} \right ) \, ,
\end{equation}
where the last equality follows from the definition of $\delta(\omega)$.
Using Eq.~\eqref{eq:app:identity}, the Einstein gradiometer time delay takes the form
\begin{equation}
\begin{aligned} \label{eq:app:td-Einstein}
\Delta \widetilde{t}_{\mathrm{grad},\mathcal{E}}^\mathrm{MZ} (\omega) %
& = \left ( e^{\mi \omega T} - 1 \right ) \left [-\frac{1}{2 \mi \, \omega} \widetilde{h}_{00}(\omega,x_{\mathrm{AI}}) e^{\mi \omega \left (x_{\mathrm{AI}} - x_\mathrm{AI_1} \right )} \right ]\Bigg|^{x_\mathrm{AI_1}}_{x_\mathrm{AI_2}} \, .
\end{aligned}
\end{equation}
Finally, noting that $\left ( e^{\mi \omega T} - 1 \right )^2 =  - 4 \, e^{\mi \omega T} \sin^2 (\omega T/2) = - \omega^2 \, T^2 \, e^{\mi \omega T} \sinc^2 (\omega T/2) $ and using Eqs.~\eqref{eq:app_grad-phase-MZ}, \eqref{eq:app:td-Doppler}, \eqref{eq:app:td-Shapiro} and \eqref{eq:app:td-Einstein}, the three %
gradiometer phase shift contributions %
take the form
\begin{equation} \label{eq:app:grad-phase-MZ-FT}
\begin{aligned}
    \Delta \widetilde{\phi}^\mathrm{MZ}_{\mathrm{grad},\mathcal{E}}(\omega)  
    =&  \omega_a \, T^2 \, \omega^2\,
    K_{\rm MZ}   (\omega) \left [- \frac{1}{2\mi \omega} \widetilde{h}_{00} (\omega,x_{\mathrm{AI}}) e^{\mi \omega (x_{\mathrm{AI}} - x_\mathrm{AI_1})} \right ]\Bigg|_{x_{\mathrm{AI}_2}}^{x_{\mathrm{AI}_1}} \, , \\
    \Delta \widetilde{\phi}^\mathrm{MZ}_{\mathrm{grad},\mathcal{D}}(\omega) 
    =& \omega_a \, T^2 \, \omega^2\,
    K_{\rm MZ}   (\omega) \left [ n_i \delta \widetilde{x}^i(\omega,x_{\mathrm{AI}})e^{\mi \omega (x_{\mathrm{AI}} - x_\mathrm{AI_1})}\right] \Bigg|_{x_{\mathrm{AI}_2}}^{x_{\mathrm{AI}_1}} \, , \\
    \Delta \widetilde{\phi}^\mathrm{MZ}_{\mathrm{grad},\mathcal{S}}(\omega) 
    =& \omega_a \, T^2 \, \omega^2\,
    K_{\rm MZ}   (\omega) 
    \left[  
    \Delta \widetilde{\mathcal{T}}_\mathcal{S}^+ (\omega,x_\mathrm{AI_1}, x_{\mathrm{AI}}) \right ]\Bigg|_{x_{\mathrm{AI}_2}}^{x_{\mathrm{AI}_1}} \, ,
\end{aligned}
\end{equation}
where we defined 
$K_\mathrm{MZ}(\omega) =  e^{\mi \omega T} \sinc^2 (\omega T/2) $. 

\subsection{Large-momentum-transfer gradiometer}\label{app:LMT-config}

The derivation of the Einstein, Doppler and Shapiro gradiometer phase shift contributions for a LMT gradiometer follows in a similar fashion. As we showed in section~\ref{sec:LMT}, the gradiometer phase shift contributions take the form
\begin{equation}\label{eq:app:grad-LMT}
\begin{aligned}
\Delta \phi^\mathrm{LMT}_{\mathrm{grad},\mathcal{E}, \mathcal{D}, \mathcal{S}}(t_0) &= \omega_a \sum_{\substack{k=0 \\ \text{even}}}^{n-2} \Bigg (\Delta t_{\mathrm{grad},\mathcal{E}, \mathcal{D}, \mathcal{S}}^{(k)} (t_0) \bigg |_{\mathrm{L}_1} + \Delta t_{\mathrm{grad},\mathcal{E},\mathcal{D}, \mathcal{S}}^{(k)} (t_0+T-(n-1)L) \bigg |_{\mathrm{L}_2} \\
& \qquad \qquad \quad  - \Delta t_{\mathrm{grad},\mathcal{E},\mathcal{D}, \mathcal{S}}^{(k)} (t_0+T) \bigg |_{\mathrm{L}_1} - \Delta t_{\mathrm{grad},\mathcal{E}, \mathcal{D}, \mathcal{S}}^{(k)} (t_0+2T-(n-1)L) \bigg |_{\mathrm{L}_2}\Bigg ) \, .
\end{aligned}
\end{equation}
Here we sum over only even $k$ because only excited states segments contribute to the observable phase shift. We quote Eq.~\eqref{eq:t_k}--\eqref{eq:grad-time delays-LMT-L1L2} from Sec.~\ref{app:LMT-config} for the expressions of $\Delta t_{\mathrm{grad},\mathcal{E,D,S}}^{(k)} (t)\big |_{\mathrm{L}_1}$ and $\Delta t_{\mathrm{grad},\mathcal{E,D,S}}^{(k)} (t)\big |_{\mathrm{L}_2}$ to start our derivation:
\begin{equation}
\begin{aligned}
\left. \Delta t_{\mathrm{grad},\mathcal{E}}^{(k)} (t)\right|_{\mathrm{L}_1,\mathrm{L}_2} 
& = -\frac{1}{2}\Bigg [\int_{t_k(t,x_\mathrm{AI})}^{t_{k+1}(t,x_\mathrm{AI})}  h_{00} (t',x_{\mathrm{AI}}) dt' \Bigg] \Bigg |^{x_{\mathrm{AI}_1}} _{x_{\mathrm{AI}_2}}\,  , \\
\left. \Delta t_{\mathrm{grad},\mathcal{D}}^{(k)} (t) \right|_{\mathrm{L}_1,\mathrm{L}_2} & = n_i \Big [{(\pm)}_{(k+1)}\delta x^i(t_{k+1}(t,x_\mathrm{AI}),x_{\mathrm{AI}}) - {(\pm)}_{(k)}\delta x^i(t_k(t,x_\mathrm{AI}),x_{\mathrm{AI}}) \Big ] \Bigg |^{x_{\mathrm{AI}_1}} _{x_{\mathrm{AI}_2}}\, , \\
        \left. \Delta t_{\mathrm{grad},\mathcal{S}}^{(k)} (t)\right|_{\mathrm{L}_1,\mathrm{L}_2} 
        & = \left[\Delta\mathcal{T}^{{(\pm)}_{(k+1)}}_\mathcal{S}\left (t+(k+1)L, x^{(k+1)}_\mathrm{L}, x_{\mathrm{AI}}\right) -\Delta\mathcal{T}^{{(\pm)}_{(k)}}_\mathcal{S}\left (t+kL, x^{(k)}_\mathrm{L}, x_{\mathrm{AI}}\right)\right] \Bigg |^{x_{\mathrm{AI}_1}} _{x_{\mathrm{AI}_2}} \, ,
\end{aligned} 
\end{equation}
where $t_k(t,x)$ is defined as
\begin{equation}
	t_k (t, x) = t+kL+{(\pm)}_{(k)}(x-x_{\mathrm{L}}^{(k)}) \, .
\end{equation}
We note that $x_\mathrm{L}^{(k)}$ is the location of the laser source that emits the $k$-th pulse, and ${(\pm)}_{(k)}$ corresponds to the direction of the $k$-th laser pulse. An ``outgoing" pulse is defined to be parallel to the AG baseline $\unit{n}$ and takes the sign ``$+$", while an ``incoming" pulse is anti-parallel to the baseline and takes the sign ``$-$". In an LMT sequence, a pair of laser pulses that define a path segment always consists of two consecutive pulses from opposite ends of the baseline. There are two possible combinations of laser pulse pairs, ``$>$" consists of an outgoing-incoming pair, and ``$<$" consists of an incoming-outgoing pair. 

For the sake of clarity, in this section, we spell out all the $x_\mathrm{L}^{(k)}$ and ${(\pm)}_{(k)}$ for different path segments in an LMT sequence, and write down explicit equations for the gradiometer time delays for the two kinds of segments. We define the gradiometer time delays for a path segment initiated by a pulse from $x_{\mathrm{L}_1}$ (i.e. a ``$>$" segment) to be $\Delta t^{(k)}_{\text{grad},\mathcal{D,S,E}} \big|_{\mathrm{L}_1}$. In this case, pulse $k$ is emitted from $x_{\mathrm{L}}^{(0)}=x_{\mathrm{L}_1}$ and is outgoing (i.e. ${(\pm)}_{(k)}=+1$), while pulse $k+1$ is emitted from $x_{\mathrm{L}}^{(k+1)}=x_{\mathrm{L}_2}$ and is incoming (i.e. ${(\pm)}_{(k+1)}=-1$). We can write down the gradiometer time delays for excited state segments started by an outgoing pulse from $x_{\mathrm{L}_1}$:
\begin{equation}\label{eq:app:grad-time delays-LMT-L1}
\begin{aligned}
\left. \Delta t_{\mathrm{grad},\mathcal{E}}^{(k)} (t)\right|_{\mathrm{L}_1} 
& = -\frac{1}{2}\Bigg [\int_{t_k(t,x_\mathrm{AI})}^{t_{k+1}(t,x_\mathrm{AI})}  h_{00} (t',x_{\mathrm{AI}}) dt' \Bigg] \Bigg |^{x_{\mathrm{AI}_1}} _{x_{\mathrm{AI}_2}}\,  , \\
\left. \Delta t_{\mathrm{grad},\mathcal{D}}^{(k)} (t) \right|_{\mathrm{L}_1} 
& = - n_i \Big [\delta x^i(t_k(t,x_\mathrm{AI}),x_{\mathrm{AI}}) + \delta x^i(t_{k+1}(t,x_\mathrm{AI}),x_{\mathrm{AI}}) \Big ] \Bigg |^{x_{\mathrm{AI}_1}} _{x_{\mathrm{AI}_2}}\, ,\\
\left. \Delta t_{\mathrm{grad},\mathcal{S}}^{(k)} (t)\right|_{\mathrm{L}_1} 
& =  
- \left[\int_{x_{\mathrm{L}_1} }^{x_{\mathrm{AI}}}\mathcal{H}_{+} (t_k(t,x'),x')dx' + \int_{x_{\mathrm{L}_2} }^{x_{\mathrm{AI}}}\mathcal{H}_{-} (t_{k+1}(t,x'),x')dx'\right] \Bigg |^{x_{\mathrm{AI}_1}} _{x_{\mathrm{AI}_2}} \, ,
\end{aligned} 
\end{equation}
where we have for even $k$:
\begin{equation}\label{eq:t_k_L1}
    \begin{split}
        t_k (t, x) &= t+kL+(x-x_{\mathrm{L}_1})\, , \quad \text{for}\quad |g\rangle\rightarrow |e\rangle \, ,\\
		t_{k+1} (t, x) &=t+(k+1)L-(x-x_{\mathrm{L}_2})\, , \quad \text{for}\quad |e\rangle\rightarrow |g\rangle \, .
	\end{split}
\end{equation}

Similarly, the gradiometer time delays for a path segment initiated by a pulse from $x_{\mathrm{L}_2}$ (i.e. a ``$<$" segment) is written as $\Delta t^{(k)}_{\text{grad},\mathcal{D,S,E}} \big|_{\mathrm{L}_2}$. In this case, pulse $k$ is emitted from $x_{\mathrm{L}}^{(k)}=x_{\mathrm{L}_2}$ and is incoming (i.e. ${(\pm)}_{(k)}=-1$), while pulse $k+1$ is emitted from $x_{\mathrm{L}}^{(k+1)}=x_{\mathrm{L}_1}$ and is outgoing (i.e. ${(\pm)}_{(k+1)}=+1$). The gradiometer time delays for excited state segments started by an incoming pulse from $x_{\mathrm{L}_2}$ can be written as:
\begin{equation}\label{eq:grad-time delays-LMT-L2}
\begin{aligned}
\left. \Delta t_{\mathrm{grad},\mathcal{E}}^{(k)} (t)\right|_{\mathrm{L}_2} 
& = -\frac{1}{2}\Bigg [\int_{t_k(t,x_\mathrm{AI})}^{t_{k+1}(t,x_\mathrm{AI})}  h_{00} (t',x_{\mathrm{AI}}) dt' \Bigg] \Bigg |^{x_{\mathrm{AI}_1}} _{x_{\mathrm{AI}_2}}\,  , \\
\left. \Delta t_{\mathrm{grad},\mathcal{D}}^{(k)} (t) \right|_{\mathrm{L}_2} 
& = n_i \Big [\delta x^i(t_k(t,x_\mathrm{AI}),x_{\mathrm{AI}}) + \delta x^i(t_{k+1}(t,x_\mathrm{AI}),x_{\mathrm{AI}}) \Big ] \Bigg |^{x_{\mathrm{AI}_1}} _{x_{\mathrm{AI}_2}}\, ,\\
\left. \Delta t_{\mathrm{grad},\mathcal{S}}^{(k)} (t)\right|_{\mathrm{L}_2} 
& =  
\left[\int_{x_{\mathrm{L}_1} }^{x_{\mathrm{AI}}}\mathcal{H}_{+} (t_{k+1}(t,x'),x')dx' + \int_{x_{\mathrm{L}_2} }^{x_{\mathrm{AI}}}\mathcal{H}_{-} (t_{k}(t,x'),x')dx'\right] \Bigg |^{x_{\mathrm{AI}_1}} _{x_{\mathrm{AI}_2}} \, ,
\end{aligned} 
\end{equation}
where we have for even $k$:
\begin{equation}\label{eq:t_k_L2}
    \begin{split}
        t_k (t, x) &= t+kL-(x-x_{\mathrm{L}_2})\, ,\quad \text{for}\quad |g\rangle\rightarrow |e\rangle \, , \\
		t_{k+1} (t, x) &=t+(k+1)L+(x-x_{\mathrm{L}_1})\, , \quad \text{for}\quad |e\rangle\rightarrow |g\rangle \, .
	\end{split}
\end{equation}

In the frequency domain, using the same procedure as before, Eq.~\eqref{eq:app:grad-LMT} can be rewritten as
\begin{equation}\label{eq:app:grad-LMT-FT}
\begin{aligned}
\Delta \widetilde{\phi}^\mathrm{LMT}_{\mathrm{grad},\mathcal{E}, \mathcal{D}, \mathcal{S}}(\omega) &= \omega_a \left ( 1-e^{\mi \omega T }\right )  \sum_{\substack{k=0 \\ \text{even}}}^{n-2}\Delta \widetilde{t}_{\mathrm{grad},\mathcal{E}, \mathcal{D}, \mathcal{S}}^{(k)} (\omega) \bigg |_{\mathrm{L}_1} \\ & \qquad \qquad \qquad  + \omega_a \, e^{\mi \omega (T-(n-1)L)}\left ( 1- e^{\mi \omega T}\right ) \sum_{\substack{k=0 \\ \text{even}}}^{n-2}\Delta \widetilde{t}_{\mathrm{grad},\mathcal{E},\mathcal{D}, \mathcal{S}}^{(k)} (\omega) \bigg |_{\mathrm{L}_2} \, .
\end{aligned}
\end{equation}
Using Eq.~\eqref{eq:app:identity} and the steps shown in Appendix~\ref{app:MZ-config}, and setting $x_{\mathrm{L}_2} = x_{\mathrm{L}_1}+L$, the Fourier transforms of the Einstein gradiometer time delays for a particular value of $k$ take the form
\begin{equation}
\begin{aligned} \label{eq:app:td-Einstein-LMT}
\left. \Delta \widetilde{t}_{\mathrm{grad},\mathcal{E}}^{(k)} (\omega)\right|_{\mathrm{L}_1} &=  e^{\mi \omega k L} \left [-\frac{1}{2 \mi \, \omega} \widetilde{h}_{00}(\omega,x_{\mathrm{AI}}) \left ( e^{\mi \omega L}e^{\mi \omega \left (x_{\mathrm{L}_2}-x_{\mathrm{AI}} \right )} - e^{\mi \omega \left (x_{\mathrm{AI}} - x_{\mathrm{L}_1} \right )}\right ) \right ]\Bigg|^{x_\mathrm{AI_1}}_{x_\mathrm{AI_2}} \, , \\
\left. \Delta \widetilde{t}_{\mathrm{grad},\mathcal{E}}^{(k)} (\omega)\right|_{\mathrm{L}_2} &= e^{\mi \omega k L} \left [-\frac{1}{2 \mi \, \omega} \widetilde{h}_{00}(\omega,x_{\mathrm{AI}})  \left ( e^{\mi \omega L}e^{\mi \omega \left (x_{\mathrm{AI}} - x_{\mathrm{L}_1} \right )} - e^{\mi \omega \left ( x_{\mathrm{L}_2} -x_{\mathrm{AI}} \right )}\right ) \right ]\Bigg|^{x_\mathrm{AI_1}}_{x_\mathrm{AI_2}} \, ,
\end{aligned}
\end{equation}
while the Doppler contributions can be expressed as
\begin{equation}
\begin{aligned} \label{eq:app:td-Doppler-LMT}
\left. \Delta \widetilde{t}_{\mathrm{grad},\mathcal{D}}^{(k)} (\omega)\right|_{\mathrm{L}_1} &=  e^{\mi \omega k L} \left [-n_i \delta \widetilde{x}^i (\omega,x_{\mathrm{AI}}) \left ( e^{\mi \omega \left (x_{\mathrm{AI}} - x_{\mathrm{L}_1} \right )} + e^{\mi \omega L}e^{\mi \omega \left (x_{\mathrm{L}_2}-x_{\mathrm{AI}} \right )}\right ) \right ]\Bigg|^{x_\mathrm{AI_1}}_{x_\mathrm{AI_2}} \, , \\
\left. \Delta \widetilde{t}_{\mathrm{grad},\mathcal{D}}^{(k)} (\omega)\right|_{\mathrm{L}_2} &=  e^{\mi \omega k L} \left [n_i \delta \widetilde{x}^i (\omega,x_{\mathrm{AI}})  \left ( e^{\mi \omega \left (x_{\mathrm{L}_2}-x_{\mathrm{AI}} \right )} + e^{\mi \omega L}e^{\mi \omega \left (x_{\mathrm{AI}} -x_{\mathrm{L}_1} \right )}\right ) \right ]\Bigg|^{x_\mathrm{AI_1}}_{x_\mathrm{AI_2}} \, .
\end{aligned}
\end{equation}
To achieve compact closed form expressions in the frequency domain, we express the Shapiro gradiometer time delay in terms of Eqs.~\eqref{eq:TScurl}. Explicitly, in terms of $x_{\mathrm{L}_1}$ and $ x_{\mathrm{L}_2}$, we find
\begin{equation} \label{eq:app:t_grad-Shapiro}
\begin{aligned}
\left. \Delta t_{\mathrm{grad},\mathcal{S}}^{(k)} (t)\right|_{\mathrm{L}_1} 
=&  
\left[\Delta \mathcal{T}^-_\mathcal{S}(t+(k+1)L, x_{\mathrm{L}_2}, x_\mathrm{AI})-\Delta \mathcal{T}^+_\mathcal{S}(t+kL, x_{\mathrm{L}_1},x_\mathrm{AI})\right] \Bigg |^{x_{\mathrm{AI}_1}}_{x_{\mathrm{AI}_2}} \, , \\
\left. \Delta t_{\mathrm{grad},\mathcal{S}}^{(k)} (t)\right|_{\mathrm{L}_2} 
=& \left[\Delta \mathcal{T}^+_\mathcal{S}(t+(k+1)L, x_{\mathrm{L}_1}, x_\mathrm{AI})-\Delta \mathcal{T}^-_\mathcal{S}(t+kL, x_{\mathrm{L}_2},x_\mathrm{AI})\right] \Bigg |^{x_{\mathrm{AI}_1}}_{x_{\mathrm{AI}_2}} \, ,
\end{aligned} 
\end{equation}
where 
\begin{equation}
\begin{aligned}
\Delta \mathcal{T}^+_\mathcal{S}(t,x_{\mathrm{L}_1},x_\mathrm{AI}) & = \int_{x_{\mathrm{L}_1}}^{x_\mathrm{AI}} \mathcal{H}_{+}(t-x_{\mathrm{L}_1}+x',x') dx' \, ,\\
\Delta \mathcal{T}^-_\mathcal{S}(t,x_{\mathrm{L}_2},x_\mathrm{AI}) & = -\int_{x_{\mathrm{L}_2}}^{x_\mathrm{AI}} \mathcal{H}_{-}(t+x_{\mathrm{L}_2}-x',x') dx' \, .
\end{aligned}
\end{equation}
In the frequency domain, Eqs.~\eqref{eq:app:t_grad-Shapiro} can be compactly expressed as
\begin{equation}\label{eq:app:t_grad-Shapiro-FT}
\begin{aligned}
\left. \Delta \widetilde{t}_{\mathrm{grad},\mathcal{S}}^{(k)} (\omega)\right|_{\mathrm{L}_1} 
=& -e^{\mi \omega kL}\left [ \Delta \widetilde{\mathcal{T}}^+_\mathcal{S} (\omega, x_{\mathrm{L}_1},x_{\mathrm{AI}}) - e^{\mi \omega L}\Delta \widetilde{\mathcal{T}}^-_\mathcal{S} (\omega, x_{\mathrm{L}_2},x_{\mathrm{AI}}) \right ]\Bigg|_{x_\mathrm{AI_2}}^{x_\mathrm{AI_1}} \, , \\
\left. \Delta \widetilde{t}_{\mathrm{grad},\mathcal{S}}^{(k)} (\omega)\right|_{\mathrm{L}_2} 
=& \, e^{\mi \omega kL}\left [ e^{\mi \omega L} \Delta \widetilde{\mathcal{T}}^+_\mathcal{S} (\omega, x_{\mathrm{L}_1},x_{\mathrm{AI}}) - \Delta \widetilde{\mathcal{T}}^-_\mathcal{S} (\omega, x_{\mathrm{L}_2},x_{\mathrm{AI}}) \right ]\Bigg|_{x_\mathrm{AI_2}}^{x_\mathrm{AI_1}} \, .
\end{aligned}
\end{equation}
Note that all of the gradiometer time delay expressions in frequency space appear with a factor of $\exp(\mi \omega k L)$, where $k$ is summed from zero to $n-2$ %
with only the even $k$'s. Since the $k$-dependence of all contributions can be isolated in this term, the gradiometer phase shifts appear with an overall factor 
\begin{equation}\label{eq:app:factor}
\sum_{\substack{k=0 \\ \text{even}}}^{n-2} e^{\mi \omega k L}   = \frac{n}{2} \, e^{\mi \omega (n-2)L/2} \, \frac{\sinc(n\omega  L/2)}{\sinc(\omega L)} \, .
\end{equation} 
Making use of Eqs.~\eqref{eq:app:grad-LMT-FT}, \eqref{eq:app:td-Einstein-LMT}, \eqref{eq:app:td-Doppler-LMT}, \eqref{eq:app:t_grad-Shapiro-FT} and   \eqref{eq:app:factor}, the gradiometer phase shift contributions in frequency space may be rewritten compactly as
\begin{equation} \label{eq:app:LMT_tot_F}
\begin{aligned}
\Delta \widetilde{\phi}^\mathrm{LMT}_{\mathrm{grad},\mathcal{E}} (\omega) =  & \frac{1}{2} k_{\rm eff} \, T^2 \,  \omega^2 \,K_{\rm MZ}   (\omega)\, 
 \left [ -\frac{1}{2 \mi \omega} \widetilde{h}_{00} (\omega,x_{\mathrm{AI}}) \left (    
 K_{\rm LMT}^+ (\omega)e^{\mi \omega (x_{\mathrm{AI}}-x_{\mathrm{L}_1})} -K_{\rm LMT}^- (\omega)  e^{\mi \omega (x_{\mathrm{L}_2}-x_{\mathrm{AI}})}\right )  
\right]\Bigg|_{x_{\mathrm{AI}_2}}^{x_{\mathrm{AI}_1}} \, , \\
\Delta \widetilde{\phi}^\mathrm{LMT}_{\mathrm{grad},\mathcal{D}} (\omega) =  & \frac{1}{2} k_{\rm eff} \, T^2 \,  \omega^2 \,K_{\rm MZ}   (\omega)\, 
 \left [n_i \delta \widetilde{x}^i (\omega, x_{\mathrm{AI}}) \left (K_{\rm LMT}^+ (\omega)   e^{\mi \omega (x_{\mathrm{AI}}-x_{\mathrm{L}_1})} +  K_{\rm LMT}^- (\omega) e^{\mi \omega (x_{\mathrm{L}_2}-x_{\mathrm{AI}})}\right )
\right]\Bigg|_{x_{\mathrm{AI}_2}}^{x_{\mathrm{AI}_1}} \, , \\
\Delta \widetilde{\phi}^\mathrm{LMT}_{\mathrm{grad},\mathcal{S}} (\omega) =  & \frac{1}{2} k_{\rm eff} \, T^2 \,  \omega^2 \,K_{\rm MZ}   (\omega)\, 
 \left [ K_{\rm LMT}^+ (\omega) \Delta \widetilde{\mathcal{T}}^+_{\mathcal{S}} (\omega, x_{\mathrm{L}_1},x_{\mathrm{AI}}) -   K_{\rm LMT}^- (\omega) \Delta \widetilde{\mathcal{T}}^-_{\mathcal{S}} (\omega, x_{\mathrm{L}_2},x_{\mathrm{AI}})  
\right]\Bigg|_{x_{\mathrm{AI}_2}}^{x_{\mathrm{AI}_1}} \, ,
\end{aligned}
\end{equation}
with $k_\mathrm{eff} = n \omega_a$ and 
\begin{equation} \label{eq:app:KLMTall}
\begin{split}
K^{\pm}_{\rm LMT}(\omega)
= \frac{\sinc \left (\dfrac{n\omega L}{2} \right)}{\sinc \left (\dfrac{\omega T}{2} \right ) \sinc ( \omega L)}
\begin{cases}
& \sinc \left (\dfrac{\omega (T-(n-2) L)}{2}\right ) \left(1-\dfrac{(n-2) L}{T}\right) 
\quad \text{for outgoing photons~} (+)  \,  ,
\\
& \sinc \left (\dfrac{\omega (T-nL)}{2} \right )\left(1-\dfrac{nL}{T}\right) 
\quad \text{for incoming photons~} (-) \,. 
\\
\end{cases}
\end{split}
\end{equation}

\section{Tools for computing the gradiometer phase shift induced by a slow-varying weak Newtonian potential}
\label{sec:geodesic}

In this appendix we provide tools for computing the single-photon gradiometer phase shift induced by a slow-varying weak Newtonian potential.
We first derive the metric sourced by a massive point-like object moving at constant velocity $\Vec{v}_s \equiv v_s^i$. %
Finally, we collect key results which are necessary for computing the gradiometer phase shift induced by such an object (e.g., expressions for the potential in the frequency-domain, geometric factors, etc.).

In the rest frame of the source, 
spacetime can be described by the Schwarzschild metric. In the weak-field limit, the Schwarzschild metric in isotropic coordinates can be written as~\cite{Weinberg:1972kfs}:
\begin{equation}
    ds^2=(-1-2\Phi)dt^2+(1-2\Phi) \,  d{x_i}d{x^i} \, ,
\end{equation}
where $\Phi(r)\equiv -GM/r=-GM/\sqrt{x_i x^i}$ is the gravitational potential sourced by the massive point-like object. %
Boosting to the laboratory frame, %
the metric becomes

\begin{equation}
    \begin{split}
        ds^2=&\left[-1-2\left(1+\frac{2v_s^2}{1-v_s^2}\Phi_{v_s}\right)\right]dt^2+\left[1-2\left(1+\frac{2v_{s}^2}{1-v_s^2}\right)\Phi_{v_s}\right]dx_idx^i \\
        &- 8 \frac{v_{s,i}}{1-v_s^2} \Phi_{v_s}dt dx^i -8\frac{v_{s,i} v_{s,j}}{1-v_s^2} \Phi_{v_s} dx^i dx^j \, ,
    \end{split}
\end{equation}
where $\Phi_{v_s}=-GM/{\left|\gamma(\Vec{r}-\Vec{v}t)-(\gamma-1)(\Vec{r}-(\Vec{r}\cdot\Vec{n})\Vec{n})\right|}$ is the boosted potential, with $\gamma = 1/\sqrt{1-v_s^2}$, and the last term is evaluated for $i\neq j$.
For a slow moving source, we can express the metric to $\mathcal{O}(v_s)$, i.e.
\begin{equation}
    ds^2=-\left(1+2\Phi\right)dt^2+\left(1-2\Phi\right) dx_i dx^i -8\Phi v_{s,i} dt dx^i 
    \label{eq:metric_Ov} \, .
\end{equation}
We note that the diagonal terms reproduce the static weak field metric in the isotropic coordinates, and the leading order relativistic corrections appears in $g_{0i}$.

We now derive the Fourier transform of the Newtonian potential due to a moving point source, which enters the Fourier-transformed signal in Eqs.~\eqref{eq:phi_MZ_New}-\eqref{eq:dphi_LMT_grad_low_omega}. The trajectory of a point mass moving with a constant velocity $\vec{v}_s$ can be written as $\vec{r}_s=\vec{b}+\vec{v}_s(t-t_s)$, where $\vec{b}$ is its impact parameter (defined with respect to the origin), and $t_s$ is the time when it is closest to the origin. The potential evalauted at $(t,\vec{x})$ is then given by
\begin{equation}\label{eqn:potential}
\Phi (t, \mathbf{x}) = -\frac{GM}{| \vec{b} + \vec{v}_s (t-t_s) - \vec{x}|} \, ,
\end{equation} 
where $M$ is the mass of the object. The Fourier transform of the potential is 
\begin{equation}\label{eqn:potentialFT}
\widetilde{\Phi} (\omega, \mathbf{x}) =  -\frac{2GM}{v_s}\, e^{-i\omega \left ( t_s + \frac{\unit{v}_s \cdot \mathbf{x}}{v_s} \right )}
K_0 \left( \frac{r_{\perp}(\vec{x})}{v_s}\, \omega \right) ,
\end{equation} 
where we have defined the transverse component $\mathbf{r}_\perp$ and its magnitude to be
\begin{align}
    \mathbf{r}_\perp(\mathbf{x}) &\equiv \mathbf{b}-\mathbf{x}+(\unit{v}_s\cdot\mathbf{x})\unit{v}_s \, ,\\
    r_{\perp} (\vec{x}) &\equiv \sqrt{ |\vec{b} - \vec{x}|^2 - (\unit{v}_s \cdot \vec{x})^2 } \, ,
\end{align} 
and $v_s\equiv |\vec{v}_s|$. The gradient of the potential, which induces accelerations on test masses, takes the following form in Fourier space
\begin{equation}\label{eqn:gradpotentialFT}
\nabla \widetilde{\Phi} (\omega, \mathbf{x}) =  -\frac{2GM}{v_s^2}\, 
\omega e^{-i\omega \left ( t_s + \frac{\unit{v}_s \cdot \mathbf{x}}{v_s} \right )} \, \left[ 
\unit{r}_\perp(\vec{x}) \,K_1 \left( \frac{r_{\perp}(\vec{x})}{v_s}\, \omega \right)
- i\, \unit{v}_s\,K_0 \left( \frac{r_{\perp}(\vec{x})}{v_s}\, \omega \right) \right].
\end{equation} 
Here, $K_0(x)$ and $K_1(x)$ are modified Bessel functions of the second kind. The argument $\mathbf{x}$ is the location of the AI that measures the phase shift. For gradiometer configurations, these functions should be evaluated at the two baseline-separated AIs, with the difference evaluated according to, e.g., Eq.~\eqref{eq:phi_MZ_New} and Eq.~\eqref{eq:point_source_phase}.
Lastly, the Shapiro phase shift is related to the integrated potential along the baseline, which is given by
\begin{equation}
\int_{-\infty}^{\infty} d t e^{-\mi \omega t} \int_{-L / 2}^{L / 2} d x^{\prime} \Phi\left(t+\frac{L}{2} \pm x^{\prime}, \mathbf{x}_{\text {mid}}+x^{\prime} \mathbf{n}\right)=-\frac{2GML}{v_s} e^{-\mi\omega ( t_s + \vec{x}_\mathrm{mid} \cdot \vec{\hat{v}_s}/v_s-L/2)}
K_0 \left( \frac{r_{\perp}(\vec{x}_\mathrm{mid})}{v}\, \omega \right) \, ,
\end{equation}
for $b \gg L$, and
\begin{equation}
\begin{aligned}
\int_{-\infty}^{\infty} d t e^{-\mi \omega t} \int_{-L / 2}^{L / 2} d x^{\prime} \Phi\left(t+\frac{L}{2} \pm x^{\prime}, \mathbf{x}_{\text {mid}}+x^{\prime} \mathbf{n}\right) & =-\frac{2 G M \pi}{ \omega \, n_\pm'} e^{-\mi\omega ( t_s -L/2)} e^{\mi \omega (\vec{b}^\pm_\perp-\vec{x}^\pm_{\mathrm{mid},\perp})\cdot  \vec{\hat{v}_{s,\perp}^\pm}/v^\pm_{s,\perp}} \\ & \qquad \qquad \qquad \qquad \times e^{- \omega \left |(\vec{b}^\pm_\perp-\vec{x}^\pm_{\mathrm{mid},\perp})\times \vec{\hat{v}_{s,\perp}^\pm} \right | /v^\pm_{s,\perp}} \, , 
\end{aligned}
\end{equation}
for $b \ll L$, where $\mathbf{n}_{ \pm}^{\prime} \equiv \mathbf{n} \mp \mathbf{v}_s$, with $n_{ \pm}^{\prime} \equiv\left|\mathbf{n}_{ \pm}^{\prime}\right|$, and $\mathbf{b}_{\perp}^{ \pm}, \mathbf{x}_{\text {mid,} \perp}^{ \pm}$and $\mathbf{v_{s, \perp}^{ \pm}}$ are defined using the rule $\mathbf{A}_{\perp}^{ \pm}=\mathbf{A}-\left(\mathbf{A} \cdot \mathbf{n}_{ \pm}^{\prime}\right) \mathbf{n}_{ \pm}^{\prime} /\left(n_{ \pm}^{\prime}\right)^2$. Here, the argument $\mathbf{x}_{\text {mid}}$ is understood as the midpoint of the two baseline-separated AIs, i.e., $\mathbf{x}_{\text {mid }}=\left(\mathbf{x}_{\mathrm{AI}_1}+\mathbf{x}_{\mathrm{AI}_2}\right) / 2$.

\end{widetext}

\bibliography{bib}
\bibliographystyle{apsrev4-2}

\end{document}